\documentclass[a4paper,UKenglish,cleveref, autoref, thm-restate]{lipics-v2021}

\pdfoutput=1 \hideLIPIcs  

\title{Near-optimal edge partitioning via intersecting families} 

\author{Alexander Yakunin}{
    T-Technologies, Moscow, Russia \and  
    Moscow Institute of Physics and Technology, Moscow, Russia
}{al.v.yakunin@gmail.com}{https://orcid.org/0009-0005-3144-1815}{}
\author{Andrey Kupavskii}{
    Laboratory of Combinatorial and Geometric Structures, Moscow Institute of Physics and Technology, Moscow, Russia \and
    T-Technologies, Moscow, Russia
}{kupavskii@ya.ru}{https://orcid.org/0000-0002-8313-9598}{}
\author{Alexander Sushin}{
    T-Technologies, Moscow, Russia
}{sushin2005@mail.ru}{https://orcid.org/0009-0003-8175-4225}{}
\author{Stanislav Moiseev}{
    T-Technologies, Moscow, Russia
}{s.moiseev@t-tech.dev}{https://orcid.org/0009-0007-0599-8999}{}

\authorrunning{A. Yakunin et al.} 
\Copyright{Alexander Yakunin, Andrey Kupavskii, Alexander Sushin, Stanislav Moiseev} 

\category{} 
\relatedversion{}

\nolinenumbers 
\usepackage{tikz}

\EventEditors{John Q. Open and Joan R. Access}
\EventNoEds{2}
\EventLongTitle{42nd Conference on Very Important Topics (CVIT 2016)}
\EventShortTitle{CVIT 2016}
\EventAcronym{CVIT}
\EventYear{2016}
\EventDate{December 24--27, 2016}
\EventLocation{Little Whinging, United Kingdom}
\EventLogo{}
\SeriesVolume{42}
\ArticleNo{23}

\begin{document}

\maketitle

\begin{abstract}

We study the problem of edge partitioning, where the goal is to partition the edge set of a graph into $k$ parts.
The replication factor of a vertex $v$ is the number of parts that contain edges incident to $v$.
The goal is to minimize the average replication factor of the vertices while keeping the sizes of the parts nearly equal.
We study the regime where the number of parts is significantly smaller than the size of the graph. 

To this end, we prove asymptotically tight bounds on the optimal replication factor, both for any constant number of parts $k$ and when $k$ grows slowly with the number of vertices.
In particular, for growing $k$, every graph admits an almost balanced partition with average replication factor $\sqrt{k}(1+o(1))$, and this bound is tight.
The upper bounds are achieved by a new class of edge partitioning algorithms.
These algorithms are computationally efficient, including in the LOCAL and CONGEST models, and can be implemented as stateless streaming algorithms in graph processing frameworks.
The lower bounds are witnessed by complete graphs and by jumbled graphs, also known as pseudo-random graphs.

Our method generalizes a family of algorithms based on symmetric intersecting families of sets.
Informally, we replace the symmetry condition by a weaker balance condition that is still sufficient for the algorithms.
This relaxation makes it possible to construct such families with asymptotically optimal rank $\sqrt{k}(1+o(1))$.

\end{abstract}

\newpage
\tableofcontents
\newpage

\newcommand{\N}{\mathbb{N}}
\newcommand{\Z}{\mathbb{Z}}
\newcommand{\Q}{\mathbb{Q}}
\newcommand{\R}{\mathbb{R}}

\newcommand{\bP}{\mathbb{P}}
\newcommand{\E}{\mathbb{E}}
\newcommand{\D}{\mathbb{D}}
\newcommand{\I}{\mathbb{I}}

\newcommand{\cF}{\mathcal{F}}
\newcommand{\cG}{\mathcal{G}}

\newcommand{\ib}{\operatorname{ib}}
\newcommand{\rf}{\operatorname{rf}}
\newcommand{\arf}{\operatorname{arf}}
\newcommand{\mrf}{\operatorname{mrf}}
\newcommand{\ar}{\operatorname{ar}}
\newcommand{\mr}{\operatorname{mr}}

\newcommand{\floor}[1]{\left \lfloor{#1}\right \rfloor}
\newcommand{\floorsqrt}[1]{\left \lfloor{\sqrt{#1}}\right \rfloor}
\newcommand{\ceil}[1]{\left \lceil{#1}\right \rceil}
\newcommand{\ceilsqrt}[1]{\left \lceil{\sqrt{#1}}\right \rceil}

\newcommand{\ErdosRenyi}{Erd\H os--R\'enyi }

\section{Introduction}
\label{s-intro}

We study the following edge partitioning problem. The input is a graph\footnote{By graphs we mean both directed and undirected graphs. For the purposes of the problem considered here, edge directions usually do not matter, and undirected graphs are a special case of directed graphs. We explicitly state when the distinction is important.} $G = (V, E)$ and an integer $k$. The goal is to partition the edge set into $k$ parts. Formally, an \textit{edge partition} (or simply a \textit{partition}) is a function $f: E \rightarrow \{1,\dots,k \}$.

We introduce two metrics that measure the quality of a partition.
\begin{definition}[Metrics of an edge partition]
    \label{d-metrics}
    Let $G = (V, E)$ be a graph, and let $f: E \rightarrow \{1,\dots,k\}$ be a partition.
    
    The \emph{imbalance} of an edge partition is the ratio of the largest part size to the average part size:
    \[
    \ib_G(f) = \frac{\max_{x \in \{1,\dots,k\}}|\{ e \in E : f(e)=x \}|}{|E| / k} .
    \]
    The \emph{replication factor} of a vertex $v$ is the number of parts that contain at least one edge incident to $v$:
    \[
    \rf_G(f, v) = | \{ f(e) \; : \; e \in E , \; e \text{ is incident to }v\} | .
    \]
    The maximum and average replication factors of a graph are, respectively, the maximum and average of the replication factors of its vertices:
    \[
        \mrf_G(f) = \max_{v \in V} \rf_G(f, v);
        \quad \arf_G(f) = \frac{1}{|V|}\sum_{v \in V} \rf_G(f, v).
    \] 
    We usually omit the subscript $G$ when it is clear from the context.
\end{definition}

The problem of finding a good edge partition is motivated by distributed
vertex-cut graph processing frameworks such as PowerGraph~\cite{gonzalez2012powergraph} and GraphX~\cite{gonzalez2014graphx}, the
latter being part of Apache Spark~\cite{Zaharia10Spark}. In these frameworks, the partition $f$
determines the allocation of edges to computational nodes. The imbalance measures how evenly the edges, and hence the computation, are distributed among the nodes. The replication factor captures the communication overhead.

Generally speaking, we would like to minimize both metrics. In practice, imbalance is usually more important: both metrics contribute to computational slowdown~\cite{Kumar2017CostModel}, but high imbalance also affects memory requirements on computational nodes.
Therefore, we seek algorithms that achieve almost perfect imbalance, $\ib(f) = 1 + o(1)$, and, subject to this constraint, minimize the replication factor. We are mainly interested in the regime where $k \ll |V|$.

An important class of edge partitioning algorithms is based on intersecting families:

\begin{definition}
    Let $X$ be a $k$-element set. A collection of subsets $\cF \subset 2^X$ is called a \emph{family of sets}, or simply a \emph{family}. A family $\cF \subset 2^X$ is called
    \begin{itemize}
        \item \emph{intersecting} if any two sets in $\cF$ have non-empty intersection;
        \item \emph{symmetric} (or \emph{transitively symmetric}) if for any two elements $x, y \in X$ there exists a bijection $\varphi: X \rightarrow X$ such that $\varphi(x) = y$ and
        \[
            \{\varphi(A) \, : \, A \in \cF \} = \cF ;
        \]
    \end{itemize}
\end{definition}

\begin{figure}[t]
    \centering
    \begin{minipage}[b]{0.4\textwidth}
        \centering
        \resizebox{\textwidth}{!}{        \begin{tikzpicture}
            \definecolor{colF1}{RGB}{190,170,225}
            \definecolor{colF2}{RGB}{250,235,140}
            \definecolor{colF3}{RGB}{190,225,245}

            \definecolor{elemBlue}{RGB}{31,119,180}
            \definecolor{elemRed}{RGB}{214,39,40}
            \definecolor{elemGreen}{RGB}{44,160,44}

            \coordinate (c1) at (90:0.9);
            \coordinate (c2) at (210:0.9);
            \coordinate (c3) at (330:0.9);

            \begin{scope}[fill opacity=0.6]
                \fill[colF1, rotate around={180:(c1)}] (c1) ellipse (2.2 and 1.2);
                \fill[colF2, rotate around={300:(c2)}] (c2) ellipse (2.2 and 1.2);
                \fill[colF3, rotate around={60:(c3)}]  (c3) ellipse (2.2 and 1.2);
            \end{scope}
            \draw[rotate around={180:(c1)}] (c1) ellipse (2.2 and 1.2);
            \draw[rotate around={300:(c2)}] (c2) ellipse (2.2 and 1.2);
            \draw[rotate around={60:(c3)}]  (c3) ellipse (2.2 and 1.2);

            \node[font=\bfseries] at (0,1.551)       {$F_1$};
            \node[font=\bfseries] at (-1.346,-0.774) {$F_2$};
            \node[font=\bfseries] at (1.346,-0.774)  {$F_3$};

            \node[circle, fill=elemBlue, draw=black, text=white,
                  minimum size=6mm, inner sep=0pt] at (1.154,0.666) {$1$};
            \node[circle, fill=elemRed, draw=black, text=white,
                  minimum size=6mm, inner sep=0pt] at (-1.154,0.666) {$2$};
            \node[circle, fill=elemGreen, draw=black, text=white,
                  minimum size=6mm, inner sep=0pt] at (0,-1.331) {$3$};
        \end{tikzpicture}        }
    \end{minipage}
    \hfill
    \begin{minipage}[b]{0.4\textwidth}
        \centering
        \includegraphics[width=\textwidth]{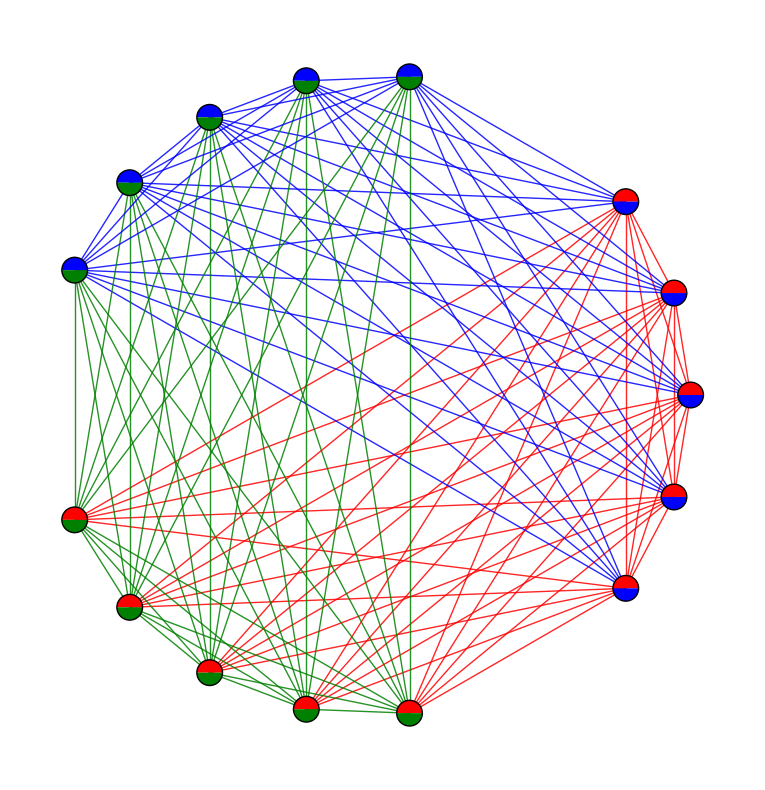}
    \end{minipage}
    \caption{A symmetric intersecting family on $3$ elements and corresponding edge partition of the complete graph. The partition is perfectly balanced and has a replication factor of 2.}
    \label{f-sim-part}
\end{figure}

Such families can be used to construct edge partitioning algorithms as follows. Let $\cF$ be a symmetric intersecting family, or SIF for short, on $k$ elements.
The algorithm assigns a random set $F_v \in \cF$ to every vertex $v$.
Then every edge $vu \in E$ is assigned to an element\footnote{The reader may think of this as choosing an element of $F_v \cap F_u$ uniformly at random. In practice, algorithms often use more sophisticated strategies for choosing an element from the intersection.}
from the intersection $F_v \cap F_u$; see Figure~\ref{f-sim-part} for an illustration.

Since the family is intersecting, the algorithm is correct: it can always assign an edge to some part. Since the family is symmetric, all parts have the same expected number of edges, and hence, using standard concentration bounds, the algorithm produces a partition with almost perfect balance. The replication factor of any vertex is at most the rank\footnote{The rank of a family $\cF$ is the maximum size of a set in $\cF$.} of the family $\cF$.

Algorithms following this scheme, such as the grid-based partitions in GraphX and PowerGraph, are widely used for several reasons. They fit into the class of \textit{stateless streaming algorithms} - the part of an edge is determined by its two endpoints alone. Stateless streaming algorithms, and streaming algorithms in general, attract considerable attention~\cite{pacaci2019experimental, Verma2017Compare, abbas2018streaming} since they can be computed efficiently in distributed systems. At the same time, SIF-based algorithms provide relatively good partition quality within this class because they guarantee a bound on the replication factor of every vertex. Finally, they are simple to implement.

Optimizing these algorithms leads to the following combinatorial problem: given $k$, find a symmetric intersecting family on $k$ elements with the minimum possible rank\footnote{One might expect that we should optimize the \textit{average} size of a set in $\cF$ rather than the \textit{maximum} size. However, for SIFs these two problems are equivalent: if a SIF contains sets of different sizes, then the subfamily consisting of all sets of minimum size is again a SIF and has smaller average and maximum set size. Thus, an optimal SIF is uniform for both objectives.}.
This is an open problem. The lower bound $\ceilsqrt{k}$ is known, and it is tight for some special values of $k$. The best known upper bound for arbitrary $k$ is $\ceil{1.1527\sqrt{k}}$. It remains open whether there exists a symmetric intersecting family of rank $\sqrt{k}(1+o(1))$~\cite{Kalai2020Symmetric}. 

In the edge partitioning literature, algorithms of this kind are known as
\textit{constraint-based partition strategies}. The applied literature usually
describes them algorithmically rather than as intersecting families, but the
underlying constructions are essentially SIF-based. Notably, the families used in practice are
weaker than the state of the art: the best rank achieved for arbitrary
$k$ is $2\sqrt{k}$.

\subsection{Main results}

In this paper, we introduce a new combinatorial structure, called a balanced intersecting system (BIS).
This structure is similar to a symmetric intersecting family, but the symmetry condition is replaced by a weaker balance condition.

We find a close connection between BISs and the edge partitioning problem.
Denote by $\ar(k)$ the optimal 
average rank (i.e.~average set size) of a balanced intersecting system on $k$ elements; a formal definition is given in Subsection~\ref{ss-bis-def}.
The following theorem shows that, for fixed $k$, the optimal average replication factor of an almost balanced partition into $k$ parts is equal to $\ar(k)$.

\begin{theorem}
    \label{t-main-fixedn}
    Let $n \rightarrow \infty$, $m \rightarrow \infty$, and let $k$ be fixed.
    Then 
    \begin{enumerate}
        \item Any graph $G$ with $n$ vertices and $m$ edges has a partition $f$ such that
        \[
            \arf(f) \le \ar(k) + o(1), \quad \ib(f) \le 1 + o(1);
        \]
        \item For any $\varepsilon>0$ and any $\alpha(n) \rightarrow 0$, there exists a graph $G$ on $n$ vertices that has no partition $f$ satisfying
        \[
            \arf(f) \le \ar(k) - \varepsilon + \alpha(n), \quad \ib(f) \le 1 + \alpha(n).
        \]
    \end{enumerate}  
\end{theorem}

The analysis of the optimal BIS rank turns out to be easier than the corresponding analysis for SIFs. We obtain the following asymptotic formula for the optimal average BIS rank.

\begin{theorem}
    \label{t-main-rank}
    $\ar(k) = \sqrt{k} (1 + o(1)).$
\end{theorem}

This gives the following result in the regime where $k$ grows sufficiently slowly with $n$.

\begin{theorem}
    \label{t-main-bign}
    Let $n \rightarrow \infty$, $m \rightarrow \infty$, $k \rightarrow \infty$, and also $k^4 \log k = o(n)$, $k^2 = o(m)$. 
    Then 
    \begin{enumerate}
        \item Any graph $G$ with $n$ vertices and $m$ edges has a partition $f$ such that
        \[
            \arf(f) \le \ar(k) + o(1) \le \sqrt{k}(1 + o(1)), \quad \ib(f) \le 1 + o(1/ k);
        \]
        \item For any $\varepsilon>0$ and any $\alpha(n) \rightarrow 0$, there exists a graph $G$ on $n$ vertices that has no partition $f$ satisfying
        \[
            \arf(f) \le \sqrt{k}(1 - \varepsilon + \alpha(n)), \quad \ib(f) \le 1 + \frac{\alpha(n)}{k}.
        \]
    \end{enumerate}
\end{theorem}

The upper bound in Theorem~\ref{t-main-fixedn} is algorithmic: we provide algorithms that compute such partitions efficiently in a variety of computational models. This also holds for the bound $\arf(f) \le \sqrt{k}(1 + o(1))$ in Theorem~\ref{t-main-bign}.

The lower bounds are in particular witnessed by complete graphs. We also prove that the same lower bounds are witnessed by a class of so-called \textit{jumbled} graphs. This is a class of pseudo-random graphs that includes dense Erd\H{o}s--R\'enyi random graphs and $(n,d,\lambda)$-graphs with $\lambda=o(d)$.

Later, in Section~\ref{s-results}, we give more precise formulations of our results, including algorithmic guarantees for the upper bounds, guarantees on the maximum replication factor, and the classes of graphs that provide the lower bounds.

\subsection{Notation}
\label{ss-defs}

Let $G = (V, E)$ be a graph, either directed or undirected. The set of all edges incident to a vertex $v$ is denoted by $E(v)$. For $A, B \subset V$, we denote by $E(A, B)$ the set of all edges between $A$ and $B$, in either direction.

The degree of a vertex $v \in V$ is denoted by $d_G(v)$. The minimum, average, and maximum degrees of a graph $G$ are denoted by $\delta(G)$, $\overline{d}(G)$, and $\Delta(G)$, respectively. We usually omit the subscript or argument $G$ when it is clear from the context.

We use $n$ for the number of vertices in a graph, $m$ for the number of edges in a graph, $k$ for either the number of parts in an edge partition or the size of the ground set of a family, and $l$ for the number of sets in a family.

For any number $k \in \N$, denote $[k]:=\{1, \dots , k \}$, and let $2^{[k]}$ stand for the set of all subsets of $[k]$.

We write $x = y \pm d$ if $x \in [y-d, y+d]$.

If $f$ is a function defined on a set $X$ and $A \subset X$, then
\[
    f(A) = \{f(x) \, : \, x \in A \}.
\]

If $B$ is a Boolean condition, then 
\[ 
\I[B] = \begin{cases}
    1, & \text{if } B \text{ is true}, \\
    0, & \text{if } B \text{ is false}.
\end{cases}
\]

We write $g(n)=o_n(h(n))$ if $\frac{g(n)}{h(n)} \xrightarrow{n \rightarrow \infty} 0$.
We omit the subscript $n$ when it is clear from the context.

We say that a sequence of random events $\{A_n \}_{n=1}^\infty$ happens with high probability, or w.h.p., if $\bP(A_n) \xrightarrow{n \rightarrow \infty} 1$.

We denote by $o_P(g(n))$ any sequence of random variables $\xi_n$ such that, for every $\varepsilon>0$, w.h.p.\ $|\xi_n| \le \varepsilon g(n)$.

We denote by $O_P(g(n))$ any sequence of random variables $\xi_n$ such that, for every $h(n)=\omega(g(n))$, w.h.p.\ $|\xi_n| \le h(n)$.

\subsection{Balanced intersecting systems}
\label{ss-bis-def}

In this subsection, we give the main combinatorial definitions used in the paper.

\begin{definition}[Weight-system]
    A triplet $S = (\cF, w, s)$ is called a \emph{weight-system}, or simply a \emph{system}, on $k$ elements if:
    \begin{enumerate}
        \item $\cF = (F_1, \dots, F_l)$ is a collection of subsets of $[k]$;
        \item $w \in [0, 1]^l$ and $\sum\limits_{i=1}^l w_i = 1$;
        \item $s \in [0, 1]^{l \times l \times k}$ and, for all $i,j \in [l]$,
        \[
            \sum\limits_{x \in [k]} s_{ijx} = 1.
        \]
    \end{enumerate}
\end{definition}

In what follows, we use the word {\it system} to refer to such triples, while the word {\it family} is reserved for set families in the usual sense.
The following definition extends the notion of an intersecting family to systems. 

\begin{definition}[Intersecting system]
    A system $(\cF, w, s)$ is \emph{intersecting} if
    \[
    \forall i, j \in [l] \; \forall x \in [k]: \quad s_{ijx} > 0 \implies x \in F_i \cap F_j.
    \]
\end{definition}

A weight-system describes the following procedure: choose two random sets $F_i,F_j \in \cF$ independently according to the distribution $w$, and then choose an element $x \in [k]$ according to the distribution $s_{ij*}$. If the system is intersecting, then this procedure always outputs an element $x \in F_i \cap F_j$. Note that if a system $(\cF,w,s)$ is intersecting, then the family $\cF$ is intersecting as well. 

\begin{definition}[Balance of a system]   
    Let $S=(\cF, w, s)$ be a system. For any $x \in [k]$, denote 
    \[
    A_x(S) =\sum_{i=1}^l \sum_{j=1}^l w_i w_j s_{ijx}.
    \]
    Informally, $A_x(S)$ is the probability that the procedure described above outputs $x$. The imbalance of a system is defined by
    \[
    \ib(S) = k \cdot \max_{x \in [k]} A_x(S).
    \]
    If $\ib(S) \le 1 + \varepsilon$, we say that $S$ is \emph{$\varepsilon$-balanced}. If $\ib(S) = 1$, we say that $S$ is \emph{balanced}.
\end{definition}

A system $S$ is called a \emph{balanced intersecting system} (BIS) if it is both intersecting and balanced.

\begin{definition}[Rank of a system]
    The \emph{maximum rank} and \emph{average rank} of a system $S=(\cF,w,s)$ are, respectively, the size of the largest set and the weighted average size of a set in $\cF$:
    \[
        \mr(S) = \max_i |F_i|;
        \quad \ar(S) = \sum_i w_i |F_i|.
    \]
    By $\ar(k)$ and $\mr(k)$ we denote the minimum\footnote{The definition of $\ar(k)$ is well posed: for every fixed family $\cF$, the function $\ar(S)$ is continuous as a function of $(w,s)$, and the set
    \[
        \{ (w,s) \; : \; (\cF,w,s) \text{ is a balanced intersecting system} \}
    \]
    is compact. Moreover, the number of possible families $\cF$ on $[k]$ is finite. Hence $\ar(S)$ attains its minimum on this set.}
    possible values of $\ar(S)$ and $\mr(S)$, respectively, over all balanced intersecting systems $S$ on $k$ elements.
\end{definition}

\subsection{Structure of the paper}

The rest of the paper is structured as follows. In Section~\ref{s-results}, we give full formulations of all our main results. The remaining sections are mostly independent, with the exception of Sections~\ref{s-main-proofs} and~\ref{s-final}. Section~\ref{s-bib-review} contains an overview of related work in two closely related areas: edge partitioning algorithms and combinatorial problems on intersecting families. In Section~\ref{s-algo}, we provide our algorithmic results: we describe additional versions of the algorithms, prove their properties, and explain how they can be used in distributed graph processing. In Section~\ref{s-lower-bound}, we prove the lower-bound theorems for graph partitioning. In Section~\ref{s-bis-rank}, we prove the upper and lower bounds on the optimal rank of BISs. In Section~\ref{s-main-proofs}, we combine the preceding results to prove Theorems~\ref{t-main-fixedn},~\ref{t-main-rank}, and~\ref{t-main-bign}. In Section~\ref{s-relation}, we prove the results concerning the relations between BISs and other combinatorial structures. Finally, in Section~\ref{s-final}, we summarize the main insights and describe possible directions for future work.

\section{Our results}
\label{s-results}

\subsection{Upper bounds on partitioning via BIS}
\label{ss-res-upper}

Here we present our algorithmic upper bounds for edge partitioning in the most general form allowed by our techniques. For simplicity, the reader may assume that all systems $S$ are balanced, that is, $\ib(S)=1$, and that the number of parts $k$ is fixed. In this case, the number of distinct sets $l$ in a BIS is also bounded by the constant $2^k$.

Balanced intersecting systems can be used to produce edge partitions similarly to symmetric intersecting families. The difference is that, instead of using the uniform distribution, we use $w$ and $s$ as the probability distributions for choosing sets and parts.

\begin{definition}[BISP algorithm]
\label{d-bisp-algo}
    \leavevmode \\
    \textbf{Input:}
    \begin{itemize}
        \item Graph $G=(V,E)$;
        \item System $S=(\cF,w,s)$ on $k$ elements, where $\cF=(F_1,\dots,F_l)$.
    \end{itemize}
    \textbf{Output:} \\
    Edge partition $f:E\rightarrow [k]$.

    \textbf{Algorithm:}
    \begin{enumerate}
        \item For every vertex $v$, choose a label $\sigma(v)\in [l]$ independently with probabilities
        \[
            \bP[\sigma(v)=i]=w_i.
        \]
        \item For every edge $vu\in E$, choose $f(vu)$ independently with probabilities
        \[
            \bP[f(vu)=x]=s_{\sigma(v)\sigma(u)x}.
        \]
    \end{enumerate}
\end{definition}

The following theorem gives the correctness guarantee for the algorithm: the metrics of the resulting partition are almost as good as the corresponding metrics of the system $S$.

\begin{theorem}
\label{t-algo}
Let $S$ be an intersecting system, and suppose that
\[
    k \log k \cdot \ib(S)= o\left( \frac{m}{\Delta(G)} \right)
    \quad \text{and} \quad
    k^2=o(n).
\]
Then the BISP algorithm returns a partition $f$ with the following properties: 
\begin{align}
    \mrf(f) &\le \mr(S); 
    \notag \\
    \arf(f) 
    &\le \ar(S) + O_P \left ( \frac{k}{\sqrt{n}} \right ) 
    = \ar(S) + o_P(1); 
    \notag \\
    \E [ \arf (f) ] &\le \ar(S); 
    \notag \\
    \ib(f) 
    &\le \ib(S) + O_P \left( \sqrt{\frac{\Delta(G) k \log k \cdot \ib(S)}{m}} \right)
    = \ib(S) + o_P(1).
    \notag 
\end{align}
\end{theorem}

When $k$ is fixed, Theorem~\ref{t-algo} does not apply to graphs with $\Delta(G)=\Omega(m)$. Thus, in this form, the algorithm effectively works only in the regime $m = \omega(n)$. 
A similar issue is observed empirically: algorithms similar to BISP may produce partitions with significant imbalance on graphs with highly skewed degree distributions~\cite{xie2014DegreeBased}.

To address this issue, we also give another BIS-based algorithm that treats high-degree vertices differently.
\begin{definition}[DBISP algorithm]
\label{d-deg-bisp-algo}
    \leavevmode \\
    \textbf{Input:} 
    \begin{itemize}
        \item Graph $G=(V,E)$;
        \item System $S=(\cF,w,s)$ on $k$ elements, where $\cF=(F_1,\dots,F_l)$;
        \item Threshold parameter $d_H\in \N$.
    \end{itemize} 
    \textbf{Output:} \\
    Edge partition $f:E\rightarrow [k]$.

    \textbf{Algorithm:}
    \begin{enumerate}
        \item For every vertex $v$ such that $d(v)<d_H$, choose a label $\sigma(v)\in [l]$ independently with probabilities
        \[
            \bP[\sigma(v)=i]=w_i.
        \]
        \item For every edge $vu\in E$, choose $f(vu)$ independently according to the following rule\footnote{Here we assume that $d(v)\le d(u)$. The case $d(u)\le d(v)$ is handled symmetrically.}:
        \[
        \bP[f(vu)=x] =
        \begin{cases}
            s_{\sigma(v)\sigma(u)x}, & d(v) \le d(u) < d_H, \\
            \sum_i w_i s_{\sigma(v)i x}, & d(v) < d_H \le d(u), \\
            1/ k, & d_H \le d(v) \le d(u).
        \end{cases}
        \]
    \end{enumerate}
\end{definition}

The DBISP algorithm works in the sparse regime, for example when $m=O(n)$, and provides a better imbalance bound. This comes at the cost of a slightly weaker bound on the average replication factor and no guarantee on the maximum replication factor.

\begin{theorem}
    \label{t-algo-deg}
    Let $S$ be an intersecting system, $k^2 = o(n)$, and let $d_H$ satisfy\footnote{Note that if $k^2 \log k \cdot \ib(S)=o(n)$ and $k \log k \cdot \ib(S)=o(m)$, then a proper integer threshold $d_H$ always exists.}
    \[
        d_H=\omega\left(\frac{mk}{n}\right)
        \quad \text{and} \quad
        d_H=o\left(\frac{m}{k\log k \cdot \ib(S)}\right).
    \]
    Then the DBISP algorithm with input system $S$ returns a partition $f$ with the following properties:
    \begin{align}
        \arf(f) 
        &\le \ar(S) + O \left( \frac{k\overline{d}}{d_H} \right)
        + O_P \left( \frac{k}{\sqrt{n}} \right)
        = \ar(S) + o_P(1);
     \notag \\
        \E[\arf(f)] 
        &\le \ar(S) + O \left( \frac{k\overline{d}}{d_H} \right)
        = \ar(S) + o(1);
    \notag \\
        \ib(f) 
        &\le \ib(S) + O_P \left( \sqrt{\frac{d_H \cdot \ib(S) \cdot k\log k}{m}} \right)
        = \ib(S) + o_P(1).
     \notag 
    \end{align}
\end{theorem}

Both algorithms are computationally simple and do not require structural analysis of the input graph. As a result, they can be implemented efficiently in several computational models. The following informal proposition summarizes these implementations; precise formulations and details for distributed graph-processing frameworks are given in Subsection~\ref{ss-impl}.
 
\begin{proposition}
    \label{p-impls}
    Both BISP and DBISP can be implemented in the following ways:
    \begin{enumerate}
        \item As randomized algorithms with running time $O(nl+mk)$, where $l$ is the number of sets in $S$.
        \item As deterministic algorithms with polynomial running time and weaker imbalance guarantees.
        \item As randomized $O(1)$-round algorithms in the LOCAL model.
        \item If $\log l + \log k=O(\log n)$, as randomized $O(1)$-round algorithms in the CONGEST model.
    \end{enumerate}
\end{proposition}

\subsection{Lower bounds}
\label{ss-res-lower}

In the previous section we showed that a system on $k$ elements can be used to build a partition of any graph into $k$ parts with the same imbalance and average/maximum replication factor. The following theorem gives the opposite transition for complete graphs. Thus, we show that a partition of a clique and an intersecting system are, in a precise sense, equivalent. 

\begin{theorem}
    \label{t-lower-complete}
    Let $f$ be a partition of the complete directed graph $K_n$ with loops\footnote{The results also hold for undirected graphs and/or graphs without loops, but with slightly weaker imbalance guarantees: $\ib(S)=\ib(f)\pm O(1/n)$.} into $k$ parts. Then there exists an intersecting system $S$ on $k$ elements such that 
    \[
        \ib(S)=\ib(f), \quad \ar(S)=\arf(f), \quad \mr(S)=\mrf(f).
    \]
\end{theorem}

We extend this lower bound to pseudo-random graphs. In particular, we use so-called jumbled graphs, introduced by Thomason~\cite{thomason1987PseudoRandom}. This class includes, for example, dense \ErdosRenyi random graphs and spectral expanders; see~\cite{krivelevich2006PseudoRandom} for more information.

\begin{definition}[Jumbled graph]
    We say that a graph $G$ is \emph{$\alpha$-jumbled} if, for every $U\subset V$,
    \[
        |E(U,U)| = p \binom{|U|}{2} \pm \alpha |U|,
    \]
    where $p=\frac{2m}{n^2}$.
\end{definition}

In the regime of fixed $k$, we show that the optimal replication factor for jumbled graphs is equal to the optimal rank of a BIS on $k$ elements.

\begin{theorem}
    \label{t-lower-mixing-fixedn}
    Let $n\rightarrow\infty$, and let $k$ be fixed. Let $\{G_n\}_{n=1}^\infty$ be a sequence of $o\left( pn\right)$-jumbled graphs\footnote{$o(pn)$-jumbled graphs are also called \emph{weakly pseudo-random}.}, where $p = \frac{2m}{n^2}$.

    \begin{itemize}
        \item There is no sequence of partitions $f_n$ of $G_n$ such that
        \[
            \ib(f_n) \le 1+o(1), \quad \mrf(f_n)<\mr(k).
        \]
        \item For any $\varepsilon>0$, there is no sequence of partitions $f_n$ such that
        \[
            \ib(f_n) \le 1+o(1), \quad \arf(f_n)\le \ar(k)-\varepsilon.
        \] 
    \end{itemize}
\end{theorem}

In the regime of growing $k$, we prove a $\sqrt k$ lower bound on the average replication factor.

\begin{theorem}
    \label{t-lower-mixing-bign}
    Let $n\rightarrow\infty$, 
    and $k=o(n^2)$. 
    Let $\{G_n\}_{n=1}^\infty$ be a sequence of $o\left(\frac{pn}{\sqrt k}\right)$-jumbled graphs, where $p = \frac{2m}{n^2}$.
    Then, for any $\varepsilon>0$, there is no sequence of partitions $f_n$ such that 
    \[
        \ib(f_n) \le 1+o(1/ k), \quad \arf(f_n)\le (1-\varepsilon)\sqrt k.
    \] 
\end{theorem}

\begin{figure}[t]
    \centering
    \begin{minipage}[c]{0.48\textwidth}
        \centering
        \resizebox{0.6\textwidth}{!}{        \begin{tikzpicture}
            \definecolor{colGrey}{RGB}{225,225,225}
            \definecolor{colBlue}{RGB}{190,215,240}
            \definecolor{colGreen}{RGB}{195,230,195}
            \definecolor{colRed}{RGB}{245,195,195}
            \definecolor{elemBlue}{RGB}{31,119,180}
            \definecolor{elemGreen}{RGB}{44,160,44}
            \definecolor{elemRed}{RGB}{214,39,40}
            \coordinate (p1) at (-2.5,0);
            \coordinate (p2) at (0,1);
            \coordinate (p3) at (0,0);
            \coordinate (p4) at (0,-1);
            \begin{scope}[fill opacity=0.6]
                \fill[colGrey] (p3) ellipse (0.8 and 1.55);
                \fill[colBlue,  rotate around={21.801:(-1.25,0.5)}]  (-1.25,0.5)  ellipse (1.896 and 0.5);
                \fill[colGreen, rotate around={0:(-1.25,0)}]         (-1.25,0)   ellipse (1.8 and 0.5);
                \fill[colRed,   rotate around={-21.801:(-1.25,-0.5)}] (-1.25,-0.5) ellipse (1.896 and 0.5);
            \end{scope}
            \draw (p3) ellipse (0.8 and 1.55);
            \draw[rotate around={21.801:(-1.25,0.5)}]  (-1.25,0.5)  ellipse (1.896 and 0.5);
            \draw[rotate around={0:(-1.25,0)}]         (-1.25,0)   ellipse (1.8 and 0.5);
            \draw[rotate around={-21.801:(-1.25,-0.5)}] (-1.25,-0.5) ellipse (1.896 and 0.5);
            \node[circle, fill=black, draw=black, text=white,
                  minimum size=6mm, inner sep=0pt] at (p1) {$1$};
            \node[circle, fill=elemBlue, draw=black, text=white,
                  minimum size=6mm, inner sep=0pt] at (p2) {$2$};
            \node[circle, fill=elemGreen, draw=black, text=white,
                  minimum size=6mm, inner sep=0pt] at (p3) {$3$};
            \node[circle, fill=elemRed, draw=black, text=white,
                  minimum size=6mm, inner sep=0pt] at (p4) {$4$};
        \end{tikzpicture}        }
        {\small
        \[
        \begin{gathered}
            \cF = (\{1,2\},\ \{1,3\},\ \{1,4\},\ \{2,3,4\});\\[2pt]
            w = \bigl(\tfrac14,\ \tfrac18,\ \tfrac14,\ \tfrac38\bigr);\\[2pt]
            s_{112}=s_{223}=s_{334}=s_{443}=1;\\[2pt]
            \text{the rest of $s$ is forced};\\[2pt]
            \ar(\cF, w, s) = 2+\tfrac38.
        \end{gathered}
        \]}
    \end{minipage}
    \hfill
    \begin{minipage}[c]{0.44\textwidth}
        \centering
        \includegraphics[width=\textwidth]{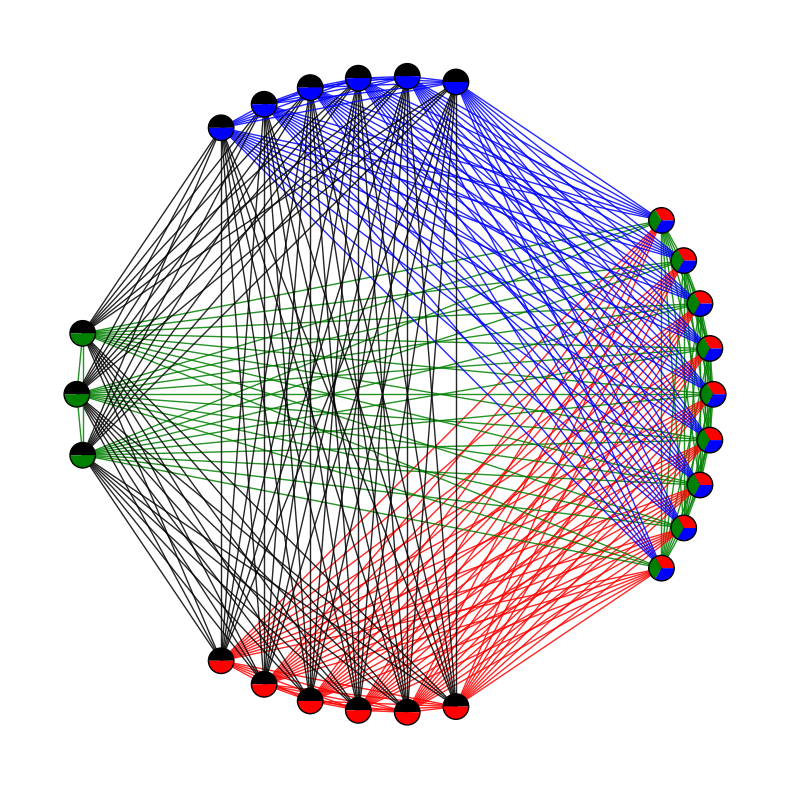}
    \end{minipage}
    \caption{An optimal BIS and corresponding partition for $k=4$.}
    \label{f-bis-n4}
\end{figure}

We know that $\ar(k) \sim \mr(k) \sim \sqrt{k}$ when $k \rightarrow \infty$ (see Corollary \ref{c-bis-rank}).
However, for a small number of parts $k$, the values $\sqrt{k}$, $\ar(k)$ and $\mr(k)$ may differ significantly. In particular, $\ar(k) > \sqrt{k}$ when $k \in \{2, 3, 4\}$ (Figures~\ref{f-bis-n2},
\ref{f-sim-part} and~\ref{f-bis-n4} show systems that are optimal for both $\ar$ and $\mr$
in these cases; their optimality can be verified via straightforward case analysis).
Thus, for small $k$ Theorems \ref{t-lower-complete} and \ref{t-lower-mixing-fixedn} provide strictly tighter bounds than the bound in Theorem \ref{t-lower-mixing-bign}. 

\subsection{Rank of BIS}
\label{ss-res-bis}

In the following theorems, we study the optimal rank of BISs and give explicit constructions.

For the upper bound, we construct a BIS of rank $\sqrt k(1+o_k(1))$. In order to be useful in a partitioning algorithm, such a BIS should also have a bounded number of sets and be efficiently computable.

\begin{theorem}
    \label{t-bis-rank-upper}
    Let
    \[
        D(r):=\min\{r-q \; : \; q\le r,\ q \text{ is a prime power}\}
    \]
    and
    \[
        r'(k)=\floor{\sqrt{k-3/4}-1/2}.
    \]
    Then there exists a balanced intersecting system $S$ on $k$ elements with 
    \[
        \mr(S) \le r'(k)+D(r'(k))+3 = \sqrt k(1+o_k(1)),
    \]
    and with at most $k$ sets. Moreover, this system can be computed in time $O(k^4)$.
\end{theorem}

We provide a matching lower bound of $\sqrt k$. For later use, we prove it more generally for slightly imbalanced intersecting systems.

\begin{theorem}
    \label{t-bis-rank-lower}
    For any $\varepsilon\ge 0$ and any $\varepsilon$-balanced intersecting system $S$, we have
    \[
        \ar(S)^2 \ge k\cdot (1-\varepsilon(k-1)).
    \]
    In particular, for any balanced system $S$, we have $\ar(S)\ge \sqrt k$.
\end{theorem}

Combining Theorems~\ref{t-bis-rank-upper} and~\ref{t-bis-rank-lower}, we obtain the following corollary.

\begin{corollary}
    \label{c-bis-rank}
    \[
        \sqrt k \le \ar(k) \le \mr(k) \le \sqrt k(1+o_k(1)).
    \]
\end{corollary}

\textbf{Proof idea for upper bound.}
The main ingredient of the upper bound is Lemma~\ref{l-system-solution-step}, which effectively shows that
\[
    \mr(k+d) \le \max(\mr(k)+1,d).
\]
To prove this lemma we take a BIS on $k$ elements, add $d$ elements to the ground set and slightly adjust the system so it is still balanced and intersecting on $k+d$ elements, at the cost of a small rank increase. In this step the balance property is crucial -- it is much harder to transform a SIF in such a manner for arbitrary $k$ and $d$, without relying on their number-theoretic properties.

We combine Lemma~\ref{l-system-solution-step} with already known examples of SIFs (that can also be turned into BIS) for good $k$: namely, projective planes for $k = q^2 + q +1$, where $q$ is a prime power. We start from the projective plane for the largest suitable prime power $q$
and apply Lemma~\ref{l-system-solution-step} repeatedly
until the ground set reaches $k$ elements. Lastly, we use the prime gap theorem to show that the rank increase is $o(\sqrt k)$.

\subsection{Relation with other intersecting family problems}
\label{ss-res-relations}

Finally, we compare BISs with other related structures studied in extremal combinatorics.

\begin{definition}
    Let $X$ be a $k$-element set. A family $\cF\subset 2^X$ is called
    \begin{itemize}
    \item \emph{regular} if all elements have the same degree, that is, lie in the same number of sets;
    \item \emph{uniform} if all sets $A\in\cF$ have the same size;
    \item \emph{balanceable} if, for some $w$ and $s$, the system $(\cF,w,s)$ is a balanced intersecting system\footnote{Note that a balanceable family is always intersecting.}.
    \end{itemize}
\end{definition}

It is immediate that symmetric families are always regular. We show that our condition on intersecting families is weaker than symmetry~\cite{Kalai2020Symmetric} and incomparable with regularity~\cite{Kupavskii2019Regular}.

\begin{theorem}
\label{t-formations-relation}
\leavevmode
\begin{enumerate}
\item Any symmetric intersecting family is balanceable.
\item There exists a uniform and balanceable family that is not regular, and hence not symmetric.
\item There exists a uniform, regular, and intersecting family that is not balanceable.
\item There exists a uniform, regular,
balanceable and intersecting family that is not symmetric.
\end{enumerate}  
\end{theorem}

\begin{figure}[t]
    \centering
    \begin{minipage}[c]{0.48\textwidth}
        \centering
        \resizebox{0.55\textwidth}{!}{        \begin{tikzpicture}
            \definecolor{colF2}{RGB}{200,195,235}
            \definecolor{colF1}{RGB}{245,205,195}
            \definecolor{elemRed}{RGB}{214,39,40}
            \definecolor{elemBlue}{RGB}{31,119,180}
            \fill[colF2] (0,0) ellipse (3.0 and 1.3);
            \draw (0,0) ellipse (3.0 and 1.3);
            \fill[colF1] (-1.2,0) circle (0.85);
            \draw (-1.2,0) circle (0.85);
            \node[circle, fill=elemRed, draw=black, text=white,
                  minimum size=7mm, inner sep=0pt] at (-1.2,0) {$1$};
            \node[circle, fill=elemBlue, draw=black, text=white,
                  minimum size=7mm, inner sep=0pt] at (1.0,0) {$2$};
        \end{tikzpicture}        }
        {\small
        \[
        \begin{gathered}
            \cF = (\{1\},\ \{1,2\});\\[2pt]
            w = \bigl(1-\tfrac{1}{\sqrt2},\ \tfrac{1}{\sqrt2}\bigr);\\[2pt]
            s_{111}=s_{121}=s_{211}=s_{222}=1;\\[2pt]
            s_{ijx}=0 \text{ for all other }i, j, x;\\[2pt]
            \ar(\cF, w, s) = 1+\tfrac{1}{\sqrt2}.
        \end{gathered}
        \]}
    \end{minipage}
    \hfill
    \begin{minipage}[c]{0.44\textwidth}
        \centering
        \includegraphics[width=\textwidth]{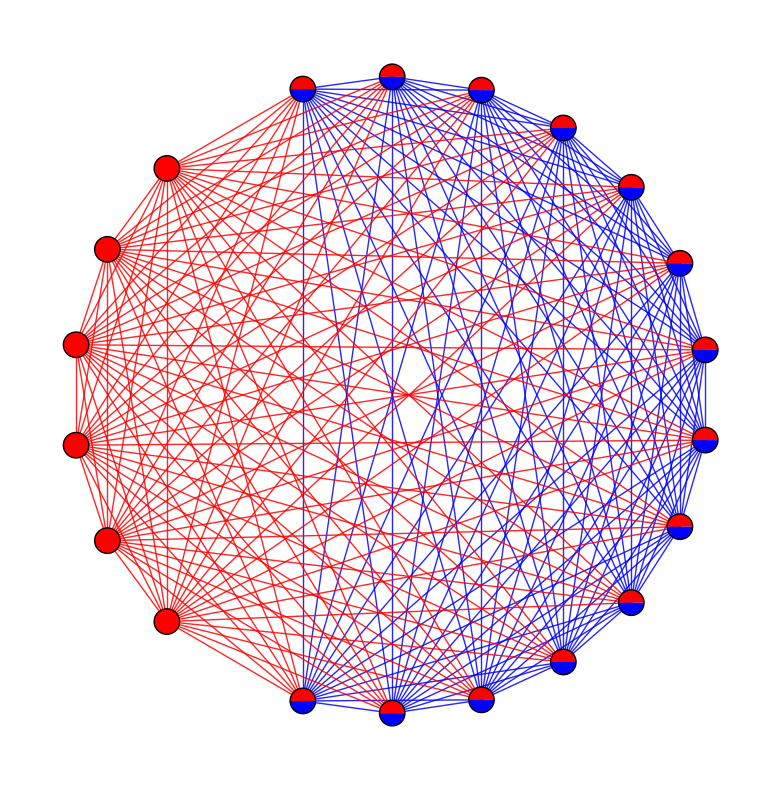}
    \end{minipage}
    \caption{An optimal BIS and corresponding partition for $k=2$.}
    \label{f-bis-n2}
\end{figure}

\section{Related work}
\label{s-bib-review}

Our work lies at the intersection of two problems: edge partitioning and minimizing the rank of intersecting systems. Here we provide an overview of both fields.

\subsection{Edge partition algorithms}
\label{ss-bib-graphs}

\textbf{Context: distributed graph frameworks.} 
Here we provide an overview of distributed graph processing frameworks. Most of them are vertex-centric: they can be seen as general algorithms that take user-defined procedures as input. These procedures describe the behavior of a vertex at each superstep: how it updates its state and what information it sends to its neighbors. Many standard graph algorithms, such as PageRank, SSSP, connected components, and K-core, can be effectively implemented in such frameworks.

Early examples of graph processing frameworks include Pregel~\cite{malewicz2010pregel} and GraphLab~\cite{low2010graphlab}. In terms of graph storage, they use the \textit{edge-cut} approach: each vertex is assigned to one computational node, and edges are duplicated on the nodes that contain their incident vertices in order to make communication between adjacent vertices possible. This allows a fairly straightforward implementation of typical graph algorithms. However, these frameworks suffer from uneven load balance when some vertices have extremely high degrees.

To address this issue, \textit{vertex-cut} frameworks were created. In these frameworks, edges are assigned to computational nodes via an edge partition, and vertices are replicated on all nodes that contain incident edges. This approach allows for better parallelization, especially on social graphs with power-law degree distributions. 

A foundational vertex-cut framework is PowerGraph~\cite{gonzalez2012powergraph}, and several other frameworks are implemented as variants of PowerGraph with additional optimizations~\cite{Li2019TopoX, Zhao2014LightGraph, chen2019PowerLyra}. GraphX~\cite{gonzalez2014graphx} is a vertex-cut framework implemented as part of Apache Spark~\cite{Zaharia10Spark}, and thus it is widely used in practice. To the best of our knowledge, it is the only such framework with a currently supported open-source implementation.
PowerLyra~\cite{chen2019PowerLyra} is notable as a so-called hybrid-cut framework, which combines the ideas of vertex-cut and edge-cut. In terms of partitioning, it uses edge partitions, but with additional constraints that allow certain algorithmic optimizations. X-Stream~\cite{Roy2013xstream} is notable as an edge-centric (but not distributed) framework: it provides a different interface to the user, where the behavior of edges must be described.

See \cite{McCune2015ReviewTLAV, Coimbra2021ReviewGraphProcessing} for more information about graph processing.

We now turn from graph-processing frameworks to the edge-partitioning algorithms used within them.

\textbf{Constraint-based algorithms.} 
The first and most known constraint-based method is grid-based, which originated from matrix computation~\cite{atalyrek2010OriginalGrid}. Roughly speaking, it is a SIF-based method where the ground set is represented as $q \times q$ square grid, and every set contains exactly one column and one row. As a result, the grid-based partition guarantees that the replication factor is at most $2 \ceilsqrt{k}$. It is implemented in many frameworks, including GraphX~\cite{gonzalez2014graphx} and PowerGraph~\cite{gonzalez2012powergraph}. The torus-based partition~\cite{jain2013GraphBuilder} improves on the grid-based partition and gives a bound of $1.5 \sqrt{k} + 1$, but it is not as widely adapted.

In their simplest form, the grid (and torus) work only when number of parts $k$ is a perfect square. This issue is not well-discussed in the literature, and different systems use different workarounds. PowerGraph restricts the grid partition to only perfect squares\footnote{\url{https://github.com/jegonzal/PowerGraph/blob/master/src/graphlab/graph/ingress/sharding_constraint.hpp}}. GraphX initially used a non-balanced strategy for non-square $k$, and later versions adopted an implementation\footnote{\url{https://github.com/apache/spark/commit/0a4071eab30db1db80f61ed2cb2e7243291183ce}} in which one column has a different size from other columns -- notably, the corresponding intersecting family is not symmetric or even regular.

PowerGraph also implemented a partitioning strategy based on perfect difference sets~\cite{Singers38OnFPP} -- it provides an optimal $\ceilsqrt{k}$ bound on replication factor for the cases where $k = q^2 + q + 1$ and $q$ is a prime.
Recent works~\cite{Qu2023LikeFPPHolistic, Zhang2023LikeFPPCombinatorial} provide perfect bounds for a broader class of $k$ -- they use projective planes and dual affine planes to achieve a bound of $\ceilsqrt{k}$ when $k = q^2 + q + 1$ or $k = q^2 + q$, and $q$ is a prime power.
They also provide a matching lower bound $\arf(f) \ge \sqrt{k}$ for a perfectly balanced partition of a complete graph.

\textbf{Other algorithms.}
Besides constraint-based methods, very few known algorithms provide theoretical guarantees on their performance.
Neighborhood Heuristic~\cite{zhang2017AlgoNe} is an algorithm that guarantees $\arf(f) \le \frac{3m + 2k +c -2}{2n}$, where $c$ is the number of connected components.
However, in the regime $\frac{m}{n} > k$ this guarantee is worse than the trivial $\arf(f) \le k$. 
Bourse, Lelarge, and Vojnovic build a polynomial-time edge-partitioning algorithm that uses vertex partitioning and prove an $O \left( \Delta(G) \sqrt{\log n \log k} \right)$ approximation factor\footnote{Note that they have slightly different definition of replication factor, satisfying $\arf'(f) = \arf(f) - 1$.} for it~\cite{bourse2014BGEP}.
These papers also prove that the problem of edge partitioning is NP-hard for the regimes of growing~\cite{bourse2014BGEP} and constant~\cite{zhang2017AlgoNe} $k$.
We next mention several notable empirical edge-partitioning algorithms.

The degree-based approach~\cite{xie2014DegreeBased} is similar to hash-based partitioning, but can also use vertex degrees. 
Despite the simplicity of distributed implementations, such algorithms can lead to poor partitioning quality, since they do not use the graph structure. As a response to this problem, greedy strategies have been actively considered, including High-Degree Replicated First (HDRF)~\cite{petroni2015hdrf}, Constell, and Zodiac~\cite{xie2014DegreeBased}. 
In these degree-based greedy approaches, edge balance is typically assigned lower priority than minimizing the replication factor.

Algorithms based on neighborhood heuristics (NE)~\cite{Hanai2019, Mayer2021, zhang2017AlgoNe}, which create partitions iteratively, lead to high partitioning quality, but there is a trade-off between partition quality and computational efficiency in such greedy approaches. Adaptive Window-based Streaming Edge Partitioning (ADWISE) from~\cite{Mayer2018Adwise} maintains information about a dynamic window of edges in order to control partitioning latency. This algorithm tries to find a balance between streaming and greedy approaches. 
JA-BE-JA-VC~\cite{Rahimian2014} is a local-search algorithm, which provides an alternative way to solve this problem. This approach can be implemented on parallel and distributed computing systems.

See \cite{Verma2017Compare} for a somewhat outdated empirical comparison of partitioning strategies.

\subsection{Intersecting families}
\label{ss-bib-family}

In this subsection, we summarize known results about two related combinatorial problems:
\begin{enumerate}
    \item Find the minimum possible rank $\rho_s(k)$ of a uniform symmetric intersecting family on $k$ elements.
    \item Find the minimum possible rank $\rho_r(k)$ of a uniform regular intersecting family on $k$ elements.
\end{enumerate}
Since any symmetric family is regular, we have $\rho_s(k) \ge \rho_r(k)$. However, known results about the two problems are very similar, so we focus on lower bounds for regular families and upper bounds for symmetric families. To the best of our knowledge, there is no known value of $k$ for which $\rho_s(k) > \rho_r(k)$.

The lower bound $\rho_r(k) \ge \sqrt k$ is known and simple. Ellis, Kalai, and Narayanan conjectured that $\rho_s(k) = \sqrt{k} (1 + o(1))$ and proved this for some specific values of $k$, mainly via geometric constructions~\cite{Kalai2020Symmetric}.

\textbf{Around finite projective planes.}
 Füredi~\cite{Furedi1981MaxDegree} proved a tight  lower bound $\rho_r(k)^2 - \rho_r(k) + 1 \ge k$. Moreover, equality holds if and only if the family is a finite projective plane:

\begin{definition}
    A \emph{finite projective plane of order $q$} is a family of sets $\mathcal{F} \subset 2^X$ such that:
    \begin{enumerate}
        \item Any two distinct sets $F, F' \in \mathcal{F}$ intersect in exactly one point, that is, $|F \cap F'| = 1$;
        \item For all distinct points $x,y \in X$, there exists exactly one set 
        $F \in \mathcal{F}$ such that $\{x,y\} \subset F$;
        \item $|X|=|\cF|=q^2+q+1$;
        \item All sets $F \in \mathcal{F}$ have size $|F| = q+1$, and every element $x \in X$ is contained in exactly $q+1$ sets.
        \footnote{Here we define projective planes as in~\cite{Furedi1981MaxDegree}. Usually properties 3 and 4 are replaced by a stronger non-degeneracy property. However, the two definitions are equivalent for order $q \ge 2$.} 
    \end{enumerate}
\end{definition}

Thus, the question about exact values of $\rho_r(k)$ when $k = r^2 - r + 1$ is harder than the question about the existence of projective planes of order $q = r-1$. When $q$ is a prime power, projective planes of order $q$ are known to exist; an example is the algebraic projective plane over the field $\mathbb{F}_q$. These planes are symmetric, so we get $\rho_s(k) = \rho_r(k) = q+1$ for $k = q^2 + q + 1$ when $q$ is a prime power. No projective plane of non-prime-power order is known. Non-existence results are also quite limited: Bruck and Ryser proved that there is no finite projective plane of order $q$ if $q$ is congruent to $1$ or $2$ modulo $4$ and $q$ cannot be expressed as a sum of two squares~\cite{Bruck1949FPPNonExistence}. The case $q=10$ was also eliminated with the help of a computational brute-force argument~\cite{Lam1989NoFPP10, Lam1991NoFPP10Exp}, and there is some progress on eliminating $q=12$~\cite{Akiyama2019TheNoFPP12}.

Any finite projective plane is a regular intersecting family by definition. A finite projective plane need not be symmetric: 
for any $q=p^{2t}$, where $p$ is odd prime and $t \in \N$, there exists a Hughes plane~\cite{Hughes1957}, which is not symmetric.

\textbf{Difference covers.} An interesting class of symmetric intersecting families comes from the so-called difference covers: let $G$ be an abelian group and let $S \subset G$. We say that $S$ is a difference cover if $S - S = G$, where $A-B$ is defined as $\{a-b \; | \; a\in A, b\in B\}$. In that case, the family $\cF = \{S+x \; | \; x \in G \}$ is a symmetric intersecting family on $|G|$ elements of rank $|S|$. In this context, cyclic groups $\Z_k$ are typically used as $G$.

Difference covers provide a simple example of a SIF of rank $\sim 2 \sqrt{k}$: we can take $S = \{1, 2, \dots, r\} \cup \{r, 2r, \dots, \floor{k /r} r \} \subset \Z$, where $r = \ceilsqrt{k}$.

An important subclass of difference covers is obtained from subsets of $\Z$: we say that $S \subset \Z$ is a difference cover for $k$ if $S - S \supset [k]$. In that case, $S$ induces a difference cover of $\Z_{2k}$ and $\Z_{2k+1}$.
This construction provides the bound $\rho_s(k) \le \ceil{1.1527 \sqrt{k}}$, which is the best known upper bound on both $\rho_r(k)$ and $\rho_s(k)$ for arbitrary $k$~\cite{Kalai2020Symmetric, Kupavskii2019Regular}. 
However, the lower bound $|S| \ge \sqrt{(2 + \frac{4}{3 \pi}) k}$ is known~\cite{RenyiBasicEstimation1949}. 
As a consequence, it is impossible to prove $\rho_s(k) = \sqrt{k} (1 + o(1))$ via difference covers in $\Z$. 

\textbf{Other problems.} Many similar problems have been studied, including finding the maximum possible number of sets in SIFs or RIFs~\cite{Cameron1989Cardinality, Kalai2020Symmetric, Kupavskii2019Regular}, $t$-wise intersecting families~\cite{Furedi1986twise}, and families that cover every pair of elements~\cite{Furedi1990CoveringPairs}.

\section{The algorithms}
\label{s-algo}

\subsection{The algorithm's correctness}
\label{ss-algo-proofs}

In this section, we prove that the partitions produced by the BISP and DBISP algorithms satisfy the properties stated in Theorems~\ref{t-algo} and~\ref{t-algo-deg}.

To avoid repetition, we use the fact that the BISP algorithm is a special case of the DBISP algorithm with $d_H=\infty$. We say that vertex $v$ is \textit{light} if $d(v) < d_H$, and \textit{heavy} otherwise. Denote by $L$ and $H$ the sets of light and heavy vertices, respectively. Recall that $S = (\cF, w, s)$ is an intersecting system, and $\cF = (F_1, \dots , F_l)$.

\textbf{Estimating replication factor.} 
Define $F(v)$ as $F_{\sigma(v)}$ if $v$ is light, and as $F(v)=[k]$ if $v$ is heavy. Since the system is intersecting, for any edge $vu \in E$ we have $f(vu) \in F(v) \cap F(u)$, and hence $f(E(v)) \subset F(v)$. Using this fact, we can estimate the replication factor of any \textit{light} vertex $v$: 
\[ 
    \rf(f, v) 
    \le |F(v)| 
    \le \mr(S);
\] \[
    \E \left[ \rf (f, v) \right]
    \le \E  [|F(v)|]
    = \sum_{i=1}^l w_i |F_i| 
    = \ar(S).
\]
Define $\arf^L(f)$ as the average replication 
factor over all \emph{light} vertices. Also define
\[
    R^L := \frac{1}{|L|}\sum_{v\in L}|F(v)|.
\]
Then $\arf^L(f)\le R^L$ and $\E [R^L]=\ar(S)$. We can estimate the deviation of $R^L$ using Chebyshev's inequality:
\[
    \D [R^L]
    = \frac{1}{|L|^2} \sum_{v \in L} \D [|F(v)|] 
    \le \frac{k^2}{|L|}
    ;
\] \[
    \forall \delta(n) = \omega \left( \frac{k}{\sqrt{|L|}} \right):
    \bP \left[ \arf^L(f) \ge \ar(S) + \delta \right]
    \le \bP \left[ R^L - \E [R^L]  \ge \delta \right]
    \le
\] \[
    \le \frac{\D [R^L]}{\delta^2}
    = o(1)
    .
\]

For the BISP algorithm, this immediately gives the desired bounds on $\mrf(f)$ and $\arf(f)$, since all vertices are light. For the DBISP algorithm, we also need to estimate the contribution of heavy vertices. Note that
the number of heavy vertices is at most $\frac{2m}{d_H}$, and the replication factor of any vertex is at most $k$. Thus we get:
\[
    \E [\arf(f)] 
    = \frac{1}{n} \sum_{v \in L} \E [\rf(f, v)] +  \frac{1}{n} \sum_{v \in H} \E [\rf(f, v)]
    \le \ar(S) + \frac{2 m k}{n d_H}
    = \ar(S) + O \left( \frac{k \overline{d}}{d_H} \right);
\] \[
    \arf(f)
    \le \arf^L(f) + \frac{1}{n} \sum_{v \in H} \rf(f, v)
    \le \ar(S) + O_P \left( \frac{k}{\sqrt{n}} \right) + O \left( \frac{k \overline{d}}{d_H} \right).
\]

\textbf{Estimating imbalance.} We start by estimating the expected size of every part.

\begin{lemma}
    For any edge $e = vu$ and any $x \in [k]$,
    \[
        \bP[f(e)=x]  \le \frac{\ib(S)}{k}.
    \]
\end{lemma}
\begin{proof}
    We need to consider three possible cases, depending on which of the vertices $v$ and $u$ are heavy. The cases are similar and straightforward; we give the details for the case $v,u \in L$.
    \[
        \bP[f(e)=x] 
        = \sum_{i, j \in [l]} \bP[\sigma(v)=i] \bP[\sigma(u)=j] \bP \left[ f(e)=x | \sigma(v)=i, \sigma(u) = j \right] = 
    \] \[
        = \sum_{i, j \in [l]} w_i w_j s_{ijx} 
        = A_x(S)
        \le \frac{\ib(S)}{k}.
        \qedhere
    \]
\end{proof}

Let $\xi_x$ denote the size of part $x$. 
Then $\xi_x = \sum_{e \in E} \I[f(e)=x]$ and $\E [\xi_x] \le \frac{\ib(S) |E|}{k}$. We now show that $\xi_x$ does not deviate significantly from its expectation.

Let $\mathcal{H}$ be a dependency graph on $\{ \I[f(e)=x] : {e \in E}\}$, that is, the vertices of $\mathcal{H}$ are indicators, and each indicator is independent from all its non-adjacent indicators;
see~\cite{Janson2004} for a formal definition. 
In our case, two indicators $\I[f(e_1)=x]$ and $\I[f(e_2)=x]$ are connected in $\mathcal{H}$ if the corresponding edges have a common \textit{light} vertex. Thus, $\Delta(\mathcal{H}) \leq D = \min(2 \Delta(G), 2 d_H)$.
We use Theorem~2.3\footnote{We state a slightly weaker form than the original Theorem 2.3. The original theorem uses more precise estimates than $\Delta(\mathcal{H})$.} from~\cite{Janson2004} to estimate deviations of $\xi_x$:

\begin{theorem}
\label{t-dependency-inequality}
Suppose that $Y_\alpha - \mathbb{E} [Y_\alpha] \leq b$ for some $b>0$ and all $\alpha \in \mathcal{A}$, and let $\mathcal{H}$ be a dependency graph on $\{Y_\alpha\}_{\alpha \in \mathcal{A}}$. Let
\[
X = \sum_{\alpha \in \mathcal{A}}  Y_\alpha, \quad V = \sum_{\alpha \in \mathcal{A}} \D [Y_\alpha].
\]
Then for $t \geq 0$:
\[\bP \left[ X \geq \mathbb{E}[X] + t \right]
  \leq \exp \left( - \frac{8t^{2}}{25 (\Delta(\mathcal{H}) + 1)(V + bt/3)} \right).\]
\end{theorem}

Let $\delta = \omega \left( \sqrt{\frac{D \cdot \ib(S)  \cdot k \log k }{m}} \right)$.
We need to prove that $\bP[\ib(f) > \ib(S) + \delta] \rightarrow 0$. 

We apply Theorem~\ref{t-dependency-inequality} to the set of variables $\mathcal{I}_x = \{ \I[f(e)=x] : {e \in E}\}$. In this case,
\[
X = \xi_x, \quad b = 1, \quad V \le \sum_{e \in E} \bP[f(e) = x] \le \frac{m \ib(S)}{k}, \quad \Delta(\mathcal{H}) \le D.
\]
Combining the union bound with Theorem~\ref{t-dependency-inequality}, we get:
\[
\bP[\ib(f) > \ib(S)+\delta] 
\le \sum_{x=1}^k \bP \left[ \xi_x > \frac{m (\ib(S) + \delta)}{k} \right]
\le \sum_{x=1}^k \bP \left[ \xi_x - \E [\xi_x] > \frac{\delta m}{k} \right]  \le
\] \[
\le k \exp \left( -\frac{8\left(\frac{\delta m}{k}\right)^{2}}{25 \left( \Delta(\mathcal{H}) + 1 \right) m\left( \frac{\ib(S)}{k} + \frac{\delta}{3k}\right)}\right)
\le 
k \exp \left(-C \cdot \frac{\delta^2 m}{Dk \ib(S)}\right),
\]
where $C$ is a constant. 
In the last inequality, we used that $\delta=o(1)=o(\ib(S))$ in the regimes of Theorems~\ref{t-algo} and~\ref{t-algo-deg}.

By the choice of $\delta$, we have $\frac{C \delta^2 m}{D \cdot k \cdot \ib(S)} = \omega(\log k)$. Therefore,
\[
    k \exp \left(-C \frac{\delta^2m}{D \cdot k \cdot \ib(S)}\right) = o(1).
\]
This concludes the proof of Theorems~\ref{t-algo} and~\ref{t-algo-deg}.

\begin{remark}[About deterministic implementation]
    \label{r-algo-imbalance-det}
    When estimating the deviation of $\xi_x$, we used the fact that all choices of $\sigma(v)$ and $f(e)$ are mutually independent. 
    We can obtain a slightly weaker estimate on the imbalance under the following weaker assumption: for any two edges $e_1 = v_1 u_1$ and $e_2 = v_2 u_2$, the variables $\sigma(v_1), \sigma(u_1), \sigma(v_2), \sigma(u_2), f(e_1), f(e_2)$ are mutually independent. This will be useful for the deterministic implementation.
    
    Under this assumption, we get:
    \[
        \D [\xi_x] = \sum_{e \in E} \D [\I[f(e)=x]]
        + \sum_{v \in L} \sum_{e_1, e_2 \in E(v)} \operatorname{cov}(\I[f(e_1)=x], \I[f(e_2)=x]) \le  
    \] \[ 
        \le \frac{\ib(S)}{k} \left( m + \sum_{v \in L} d(v)^2 \right)
        = O \left( \frac{m D \ib(S)}{k} \right).
    \]
    Hence, for any $\delta = \omega \left( \sqrt{\frac{D \cdot k^2 \cdot \ib(S)}{m}} \right)$, Chebyshev's inequality gives:
    \[
        \bP[\ib(f) > \ib(S)+\delta] 
        \le \sum_{x=1}^k \bP \left[ \xi_x > \frac{m (\ib(S) + \delta)}{k} \right]
        \le k \max_{x\in[k]}\bP \left[ \xi_x - \E [\xi_x] > \frac{\delta m}{k} \right]  \le
    \] \[
        \le  \frac{k^3 \max_{x\in[k]}\D [\xi_x]}{\delta^2 m^2} 
        = O \left( \frac{k^2 D \ib(S)}{\delta^2 m} \right) = o(1).
    \]
    Thus we get $\ib(f) \le \ib(S) + O_P\left( \sqrt{\frac{D \cdot k^2 \cdot \ib(S)}{m}} \right)$.
\end{remark}

\begin{remark}
    The DBISP algorithm works for sparse graphs, but does not give any guarantees on the maximum replication factor. It is impossible to provide such guarantees because some sparse graphs do not have any partitions with both good $\mrf$ and good $\ib$.

    Consider a star graph $G = (V, E)$, i.e.~a tree on $n$ vertices in which all vertices are connected to the center vertex $v_0 \in V$. Also consider any partition $f: E \rightarrow [k]$. Because all edges are incident to $v_0$, there are at most $\rf(f, v_0)$ non-empty parts. Thus, $\ib(f) \ge k / \mrf(f)$. In particular, any partition with $\ib(f) < 1 + \frac{1}{k-1}$ must have $\mrf(f) = k$. 
\end{remark}

\subsection{Possible implementations}
\label{ss-impl}

Here, we describe possible implementations of the BISP and DBISP algorithms, as stated in Proposition~\ref{p-impls}:

\begin{enumerate}
    \item \textbf{Randomized algorithms.}
    The randomized algorithms are straightforward implementations of BISP and DBISP.
    
    \item \textbf{Deterministic algorithms.}
    First, let us describe the deterministic guarantees of our algorithms. 
    In the deterministic version, we replace probabilistic error terms of the form $O_P(\delta(n))$ with deterministic bounds of the form $O(\delta(n))$. However, the imbalance bounds are slightly weaker than in the randomized version. In particular, for the BISP algorithm we get:
    \[ 
        \arf(f) \le \ar(S) + O \left( k / \sqrt{n}\right)
        ; \quad 
        \ib(f) \le \ib(S) + O \left( \sqrt{\frac{\Delta(G) \cdot k^2 \cdot \ib(S) }{m}} \right)
    \]
    and for DBISP:
    \[ 
        \arf(f) \le \ar(S) + O \left( \frac{k \overline{d}}{d_H} \right) + O \left( \frac{k}{\sqrt{n}} \right)
        ; \quad 
        \ib(f) \le \ib(S) + O \left( \sqrt{\frac{d_H \cdot k^2 \cdot \ib(S)  }{m}} \right)
        .
    \]
    We derandomize BISP and DBISP via hashing. Note that we do not need full independence of all random variables $\sigma(v)$ and $f(e)$. It is enough that, for any two edges $e_1,e_2\in E$, their parts and the labels of their endpoints are generated independently; here we use Remark~\ref{r-algo-imbalance-det}. Thus, instead of fully random bits, we can use a $6$-independent family of hash functions~\cite{Wegman1981KIndHash}. Since the size of such a family is polynomial, we can brute-force all such hash functions to find a suitable partition.
    
    \item \textbf{LOCAL algorithms.}
    In the LOCAL model, every vertex is treated as a machine with a unique id, and the computation proceeds in rounds. In each round, a vertex can send a message to each of its neighbors. At the end of the computation, every vertex must know the values $f(e)$ for all incident edges.

    BISP can be implemented in LOCAL as follows: every vertex generates its label $\sigma(v)$ and sends it to its neighbors in the first communication round. Then, for every edge $vu$, the source\footnote{If the graph is undirected, we say that the vertex with lower id is source vertex} vertex $v$ generates its part $f(vu)$ and sends this information to the destination vertex $u$.
    
    The described algorithm works in $2$ rounds of communication. It can be improved to work in $1$ round: before the communication, for every edge $vu$, the source vertex $v$ decides what $f(vu)$ will be for every possible value of $\sigma(u)$, and sends this information to $u$, in addition to its label $\sigma(v)$. After the communication, both the source and the destination of the edge $vu$ can compute $f(vu)$ consistently.

    DBISP is implemented very similarly. However, the DBISP algorithm also requires all vertices to know the number of edges $m$, or a constant-factor approximation of $m$; this is necessary for choosing the threshold $d_H$. It also requires one additional initial round, in which every vertex $v$ collects the degrees of all its neighbors.

    \item \textbf{CONGEST algorithms.}
    In the CONGEST model, computation proceeds as in the LOCAL model, but with the additional restriction that every vertex can send at most $O(\log n)$ bits to each neighbor in each round.

    In the LOCAL algorithms described above, the message sizes are $O(\log l + \log k)$ for the $2$-round version and $O(\log l + l \log k)$ for the $1$-round version. 
    Thus, these algorithms are valid CONGEST algorithms whenever the corresponding message size is $O(\log n)$. Note that degree collection in the DBISP algorithm requires only $O(\log n)$-bit messages.
\end{enumerate}

The described implementations cover the most common models of computation; however, they are still somewhat far from actual distributed graph processing.  
A more important property for practical applications is that both algorithms can be implemented as \emph{stateless streaming algorithms} in frameworks such as GraphX~\cite{gonzalez2014graphx}.
By a stateless streaming algorithm, we mean an algorithm that processes each edge separately, without using information about other edges, and only uses the ids of its endpoints. 

This is the preferred type of partitioning algorithm because it allows the partition to be computed in a fully parallel and distributed manner, for example while the graph is being loaded from an external source, and minimizes communication and memory overhead. Another useful property of such algorithms is that they allow one to compute the part of any edge during the algorithm using only the ids of its endpoints. 

To implement BISP as a stateless streaming algorithm, we need to assign vertex labels $\sigma(v)$ using a hash function instead of choosing them uniformly at random. To keep the theoretical guarantees, we can use $6$-independent hash functions, as in the derandomized implementation. However, in practice, even simple hash functions usually work well for similar algorithms~\cite{jain2013GraphBuilder, Qu2023LikeFPPHolistic}.

The DBISP algorithm is more costly than the BISP algorithm because it requires knowledge of vertex degrees. This means that, before partitioning the graph, we need to precompute the degrees of all vertices. After that, we only need to remember which vertices are heavy, i.e.~$O(n \overline{d}/ d_H)$ numbers. In many cases, this amount of information is small enough to be broadcast to all computational nodes.

\section{Lower bounds}
\label{s-lower-bound}

In this section, we prove the lower bounds results: Theorems~\ref{t-lower-complete},~\ref{t-lower-mixing-fixedn}, and~\ref{t-lower-mixing-bign}.
For Theorems~\ref{t-lower-complete} and~\ref{t-lower-mixing-fixedn}, we use a common construction that produces an intersecting system from a graph partition, provided the graph is sufficiently well behaved. If the graph is complete, the resulting system is intersecting and has the same parameters as the partition. If the graph is jumbled, the same properties hold up to a small error.

\subsection{Partition induces a system}\label{sec31}

Here we describe how to build a system from an edge partition.

Let $G = (V, E)$ be a graph, and let $f: E \rightarrow [k]$ be a partition. 
Let $\cF = (F_1, \dots, F_l )$ be the collection of distinct sets in $\{f(E(v)) \; |  \; v \in V\}$. Define:
\[
    V_i = \{ v \in V : f(E(v)) = F_i \};
\] \[
    E_{ij} = E(V_i, V_j), 
    \quad E_x = f^{-1}(x),
    \quad E_{ijx} = E_x \cap E_{ij};
\] \[
    w_i = \frac{|V_i|}{n},
    \quad s_{ijx} = \frac{|E_{ijx}|}{|E_{ij}|}.
\]
If $|E_{ij}| = 0$, we say that $s_{ijx}$ is undefined. Thus, for an arbitrary graph $G$, the resulting object $S = (\cF, w, s)$ is not necessarily a system. However, we can show that, whenever $s_{ijx}$ is defined, then $F_i\cap F_j\ne \emptyset$, and, moreover, the ranks of the resulting object coincide with the replication factors of $f$:

\begin{lemma}
\label{l-partition-to-system-universal} The following holds. 
\begin{enumerate}
    \item $\sum_i w_i = 1$.
    \item If there is at least one edge from $V_i$ to $V_j$, then $F_i\cap F_j\ne \emptyset$ and
    \[
        \sum_{x \in [k]} s_{ijx} = 1,
    \] \[
        \forall x \in [k]: (s_{ijx} > 0 \implies x \in F_i \cap F_j).
    \]
    \item $\ar(S) = \arf(f)$;
    \item $\mr(S) = \mrf(f)$.
\end{enumerate}
\end{lemma}

\begin{proof}
    Property 1 follows from the fact that $V = V_1 \sqcup \dots \sqcup V_l$.
    Properties 3 and 4 follow from the observation that if $v \in V_i$, then $\rf(f, v) = |F_i|$. Now let us prove the second property:
    \[
    \sum_x s_{ijx}
    = \frac{1}{|E_{ij}|} \sum_x | E_{ijx} |
    = \frac{|E_{ij}|}{|E_{ij}|} = 1.
    \]
    If $s_{ijx} > 0$, then there exists an edge $vu \in E_{ij}$ such that $f(vu)=x$. Hence
    \[
        x=f(vu)\in f(E(v))\cap f(E(u))=F_i\cap F_j.
        \qedhere
    \]
\end{proof}

Equipped with Lemma~\ref{l-partition-to-system-universal}, it is easy to finish the proof of Theorem~\ref{t-lower-complete}. We only need to prove that, in the case of a complete graph $G=K_n$, the following additional properties of the construction hold.

\begin{lemma}
\label{l-partition-to-system-complete}
Let $G = K_n$.
Then $S$ is an intersecting system and $\ib(S) = \ib(f)$.
\end{lemma}

\begin{proof}
By Lemma~\ref{l-partition-to-system-universal}, and because there are edges between every pair $V_i,V_j$, the object $S$ is an intersecting system.

Now let us show that $\ib(S) = \ib(f)$. Denote $I_v^i = \I[v \in V_i]$ and $I_{vu}^x = \I[vu \in E_x]$.

We need to calculate the value $w_i w_j s_{ijx}$. When $i = j$, we use the fact that $E_{ij} = V_i \times V_j$ to get:
\[
w_i w_j s_{ijx}
= \frac{|V_i| \cdot |V_j| \cdot |E_{ijx}|}{n^2 \cdot |E_{ij}|}
= \frac{1}{n^2} \sum_{vu \in V^2} I_v^i \cdot I_u^j \cdot I_{vu}^x
.\]
The case $i \neq j$ is slightly different since $E_{ij} = V_i \times V_j \sqcup V_j \times V_i$. Thus,
\[
  w_i w_j s_{ijx}
= \frac{|V_i| \cdot |V_j| \cdot |E_{ijx}|}{n^2 \cdot |E_{ij}|}
= \frac{1}{2 n^2} \sum_{vu \in V^2}  \bigl( I_v^i \cdot I_u^j + I_v^j \cdot I_u^i  \bigr) \cdot I_{vu}^x
.  
\]
In the sum defining $A_x(S)$, every term $I_v^i \cdot I_u^j \cdot I_{vu}^x$ appears either once with coefficient $\frac{1}{n^2}$ or twice with coefficient $\frac{1}{2n^2}$.
Thus, for any $x \in [k]$, we get
\[
A_x(S)
= \frac{1}{n^2} \sum_{i=1}^l \sum_{j=1}^l 
\sum_{(v, u) \in V^2} I_v^i \cdot I_u^j \cdot I_{vu}^x
= \frac{1}{n^2} \sum_{(v, u) \in V^2} I_{vu}^x \left( \sum_{i=1}^l \sum_{j=1}^l I_v^i \cdot I_u^j \right)
=
\] \[
= \frac{1}{n^2} \sum_{(v, u) \in V^2} I_{vu}^x
= \frac{|f^{-1}(x)|}{n^2};
\] \[
\ib(S) 
= \frac{\max_x |f^{-1}(x)|}{n^2 / k} 
= \ib(f).
\qedhere
\]
\end{proof}

This concludes the proof of Theorem~\ref{t-lower-complete}.

\subsection{Property of jumbled graphs}

We slightly reformulate the jumbledness property in a form that will be used in Theorems~\ref{t-lower-mixing-fixedn} and~\ref{t-lower-mixing-bign}. 

\begin{lemma}
    \label{l-jumbled-to-mixing}
    Let $n\rightarrow \infty$, $p = 2m / n^2$ and let $\{G_n\}$ be an $o(\delta n p)$-jumbled sequence of graphs for some $\delta(n) = \omega(1 / n)$.
    \begin{enumerate}
        \item For any $A \subset V$ such that $|A| \ge \delta n$, we have
        \[
            |E(A, A)| = \frac{p |A|^2}{2} (1 + o(1));
        \]
        \item For any disjoint $A, B \subset V$ such that $|A| \ge \delta n$ and $|B| \ge \delta n$, we have
        \[
            |E(A, B)| = p |A| |B| (1 + o(1)).
        \]
    \end{enumerate}
\end{lemma}

\begin{proof}
    Since $|A| = \omega(1)$, we have $\binom{|A|}{2} = \frac{|A|^2}{2} (1 + o(1))$. 
    
    For the first property, we estimate the deviation directly from the definition of jumbledness:
    \[
        \left|\frac{ |E(A, A)| - p |A|^2/2}{p |A|^2 /2}\right|
        \le o \left(  \frac{ \delta n p |A| }{p |A|^2} \right)
        = o \left( \frac{\delta n}{|A|} \right)
        = o(1).
    \]
    For the second property, we use the identities
    \[
        |E(A, B)| = |E(A \cup B, A \cup B)| - |E(A, A)| - |E(B, B)|;
    \] \[
        p |A| |B| = \frac{p(|A|+|B|)^2}{2} - \frac{p |A|^2}{2} - \frac{p |B|^2}{2}.
    \]
    Thus, using the jumbledness property, we get:
    \[
        \left|\frac{|E(A, B)| - p |A| |B|}{p|A||B|}\right|
        = o \left( \frac{\delta n p (|A| + |B|)}{p |A||B|} \right)
        = o \left( \frac{\delta n }{\min(|A|, |B|)} \right)
        = o(1). \qedhere
    \]
\end{proof}

\begin{remark}
    In Lemma~\ref{l-jumbled-to-mixing} and, consequently, in Theorems~\ref{t-lower-mixing-fixedn} and~\ref{t-lower-mixing-bign}, we can replace the jumbledness condition by the slightly weaker assumption that 
    \[
        |E(A,A)|=\frac{p|A|^2}{2}\pm o(\delta np|A|)
    \]
    holds for all sets $A$ such that $|A|\ge \delta n$.
\end{remark}

\subsection{Proof for constant k}

Here we prove Theorem~\ref{t-lower-mixing-fixedn}. The main part of the proof is the following lemma, which is similar to Theorem~\ref{t-lower-complete}, but applies to jumbled graphs and allows small error terms.

\begin{lemma}
    \label{l-jubmbled-part-to-bis}
    Let $n \rightarrow \infty$, let $k$ be fixed, and let $\{G_n\}$ be an $o(np)$-jumbled sequence of graphs. 
    For any sequence of partitions $f_n$ of $G_n$, there exists a sequence of intersecting systems $S_n$ on $k$ elements such that 
    \begin{itemize}
        \item $\mr(S_n) \le \mrf(f_n)$;
        \item $\ar(S_n) \le \arf(f_n) + o(1)$;
        \item $\ib(S_n) \le \ib(f_n) + o(1)$.
    \end{itemize}
\end{lemma}

\begin{proof}
    Define $\alpha(n)$ so that $G_n$ is $\alpha np$-jumbled. By assumption, we have $\alpha(n) = o(1)$. We apply Lemma~\ref{l-jumbled-to-mixing} with $\delta = \sqrt{\alpha}$.

    Let $S' = (\cF', w', s')$ be constructed from the partition $f_n$ as in Section~\ref{sec31}. The object $S'$ is not necessarily an intersecting system; we will fix this by removing all sets $F_i'$ such that $w'_i \le \delta$.

    Formally, without loss of generality, suppose that the weights $w'_i$ are sorted in decreasing order. Let $l$ be the largest index such that $w'_l > \delta$. We construct a new system $S = (\cF, w, s)$ as follows: $\cF = (F_1 \dots F_l)$ consists of the first $l$ sets of $\cF'$; $w_i = \frac{w'_i}{w'_1 + \dots + w'_l}$; and $s_{ijx} = s'_{ijx}$.

    For every $i \le l$, the set $V_i$ has at least $\delta n$ vertices. 
    By Lemma~\ref{l-jumbled-to-mixing}, for every $i,j\le l$, the quantity $|E(V_i,V_j)|$ is positive for all sufficiently large $n$.
    Hence, by the construction of $\cF'$, the system $S$ is intersecting.
    
    The total weight of the removed sets is at most $\delta 2^k = o(1)$, and hence $w_i \le \frac{w'_i}{1 - \delta 2^k}$. Now we analyze the balance and the rank of $S$:
    \[
        \mr(S) \le \mr(S') = \mrf (f_n);
    \] \[
        \ar(S) 
        = \sum_{i=1}^l w_i |F_i| 
        \le \frac{1}{1 - \delta 2^k} \sum_{i=1}^l w'_i |F_i|  
        \le \arf(f_n) (1 + o(1)).
    \]
    By Lemma~\ref{l-jumbled-to-mixing}, for $i\neq j$ we have
    \[
        |E_{ij}| \ge (1-o(1))p|V_i||V_j|,
    \]
    while for $i=j$ we have
    \[
        |E_{ii}| \ge (1-o(1))\frac{p|V_i|^2}{2}.
    \]

    Therefore, for all $i,j\le l$ and all $x\in[k]$,
    \[
        A_x(S) 
        \le \frac{1}{(1 - \delta 2^k)^2} \sum_{i, j \in [l]} \frac{|V_i| |V_j| |E_{ijx}|}{n^2 |E_{ij}|} =
    \] \[
        = \frac{1 + o(1)}{n^2 p} \left( 
            \sum_{i \in [l]} 2 |E_{iix}| + \sum_{i \neq j} |E_{ijx}|
        \right)
        = \frac{2 |E_x| (1 + o(1))}{n^2  p} = \frac{|E_x| (1 + o(1))}{m}.
    \]
    In the last transition we used that $E_{ijx} = E_{jix}$ and 
    \[
        E_x = \bigsqcup_{i \le j} E_{ijx}.
    \] 
    Finally, we get
    \[
        \ib(S)
        = k \max_x A_x(S) 
        \le k \max_x \frac{|E_x| (1 + o(1))}{m} 
        \le \ib(f) + o(1).
        \qedhere
    \]
\end{proof}

Lemma~\ref{l-jubmbled-part-to-bis} shows that if we have good partitions of jumbled graphs, then we have good intersecting systems with arbitrarily small imbalance error, that is, with imbalance arbitrarily close to $1$. The following lemma is needed to pass from arbitrarily small imbalance error to perfect balance.

\begin{lemma}
    \label{l-system-limits}
    Let $\{ S_n \}_{n=1}^\infty$ be a sequence of intersecting systems on $k$ elements, with no repeated sets in the underlying families. Then there exists an intersecting system $S^*$ such that
    \[
        \ib(S^*) \le \limsup_{n\to\infty} \ib(S_n), \quad 
        \ar(S^*) \le \limsup_{n\to\infty} \ar(S_n), \quad 
        \mr(S^*) \le \limsup_{n\to\infty} \mr(S_n).
    \]
\end{lemma}
\begin{proof}

    Since there are only finitely many possible families $\cF \subset 2^{[k]}$ without repeated sets, there exists a family $\cF^*$ that occurs infinitely many times in the sequence $\{ \cF_n \}_{n=1}^\infty$. Passing to this subsequence, the sequences $\{ w^{(n)} \}_{n=1}^\infty$ and $\{ s^{(n)} \}_{n=1}^\infty$ are bounded, and therefore have convergent subsequences.
    
    Let $\{ n_r \}_{r=1}^\infty$ be a subsequence such that $\cF^{(n_r)} = \cF^*$ for all $r$, and such that both sequences $\{ w^{(n_r)} \}_{r=1}^\infty$ and $\{ s^{(n_r)} \}_{r=1}^\infty$ converge. Let $w^*$ and $s^*$ be their limits.

    Then $S^* = (\cF^*, w^*, s^*)$ is an intersecting system, because the set of pairs $(w,s)$ for which $(\cF,w,s)$ is an intersecting system is closed. Since $\ib$, $\ar$, and $\mr$ are continuous on this fixed family, we obtain the desired inequalities.
\end{proof}

Now let us prove Theorem~\ref{t-lower-mixing-fixedn} by contradiction. Suppose that $\{G_n\}_{n=1}^\infty$ is a given sequence of $o(np)$-jumbled graphs, and that $\{f_n\}_{n=1}^\infty$ is a sequence of partitions such that $\ib(f_n) \le 1 + o(1)$ and either $\mrf(f_n) < \mr(k)$ for all $n$, or $\arf(f_n) \le \ar(k) - \varepsilon$ for all $n$. By Lemma~\ref{l-jubmbled-part-to-bis}, we get a sequence of intersecting systems $\{S_n\}_{n=1}^\infty$ with the corresponding bounds on imbalance and rank. By Lemma~\ref{l-system-limits}, we get a balanced intersecting system $S^*$ with either $\mr(S^*) < \mr(k)$ or $\ar(S^*) \le \ar(k) - \varepsilon$. This is impossible by the definitions of $\ar(k)$ and $\mr(k)$, a contradiction. This proves Theorem~\ref{t-lower-mixing-fixedn}.

\subsection{Proof for growing k}

\textbf{Idea of the proof:} define $E_x = f^{-1}(x)$ and let $V_x$ be the set of all vertices incident to at least one edge of $E_x$. Then $E_x \subset E(V_x, V_x)$. If $f$ is balanced and $G$ is jumbled, then
\[
    \frac{m}{k}
    \approx |E_x| 
    \lesssim \frac{p |V_x|^2}{2};
\] \[
    |V_x|
    \gtrsim \sqrt{\frac{2 m}{k p}} 
    \approx \frac{n}{\sqrt{k}};
\] \[
    \arf(f) 
    = \frac{1}{n} \sum_{x=1}^k |V_x| 
    \ge \sqrt{k}.
\]
\textbf{Formal proof:} suppose that $G$ is an $o(np/\sqrt{k})$-jumbled graph and that $f$ is a partition of $G$ with $\ib(f) \le 1 + \alpha$, where $\alpha k=o(1)$. We will show that $\arf(f)\ge \sqrt k(1+o(1))$. Define $\delta = \frac{1}{2 \sqrt{k}}$. We can apply Lemma~\ref{l-jumbled-to-mixing} to $G$ with this value of $\delta$.

Define $E_x = f^{-1}(x)$. 
We know that $|E_x| \le (1 + \alpha) m / k$ for every $x$ and that $\sum_x |E_x| = m$, so
\[
    |E_x|
    = m - \sum_{y \neq x} |E_y|
    \ge m - (k-1)\frac{(1+\alpha)m}{k}
    \ge \frac{m (1 - \alpha k)}{k}.
\]
Define $V_x = \{v: E(v) \cap E_x \neq \varnothing \}$.  
Define $V_x'$ as follows: start with $V_x' = V_x$; if $|V_x'| < \delta n$, add arbitrary vertices to $V_x'$ until $|V_x'| = \ceil{\delta n}$.

Note that $E_x \subset E(V_x, V_x) \subset E(V_x', V_x')$ and $|V_x'| \ge \delta n$. By Lemma~\ref{l-jumbled-to-mixing},
\[
    |E_x| \le \frac{p}{2} |V'_x|^2 (1 + o(1)).
\]
Combining these bounds on $|E_x|$, we get
\[
    \frac{p}{2} |V'_x|^2 (1 + o(1)) \ge \frac{m (1 - \alpha k)}{k};
\] \[
    |V'_x| \ge \frac{n}{\sqrt{k}} (1 + o(1)),
\]
where we used $p=2m/n^2$ and $\alpha k=o(1)$.
Since $k=o(n^2)$, we have $n/\sqrt k\to\infty$, and hence, for all sufficiently large $n$,
\[
    \ceil{\delta n} 
    \le \frac{n}{2 \sqrt{k}} +1 
    < |V'_x|.
\] 
Thus $V_x' = V_x$. Finally, we estimate
\[
    \arf(f)
    = \frac{1}{n} \sum_{v \in V} \sum_{x=1}^k \I [v \in V_x]
    = \frac{1}{n} \sum_x |V_x| 
    \ge \sqrt{k} (1 + o(1)).
\]
This contradicts the existence of a sequence of partitions with $\ib(f_n)\le 1+o(1/ k)$ and $\arf(f_n)\le (1-\varepsilon)\sqrt k$. Therefore, Theorem~\ref{t-lower-mixing-bign} follows.

\section{Rank of BIS}
\label{s-bis-rank}

In this section, we prove our estimates on the optimal BIS rank. We start with the simpler lower bound, namely Theorem~\ref{t-bis-rank-lower}. Then we prove the upper bound, namely Theorem~\ref{t-bis-rank-upper}.
\subsection{Lower bound}

Let $S =(\cF, w, s)$ be an $\varepsilon$-balanced intersecting system. Here we write $A_x$ instead of $A_x(S)$ for simplicity.

Recall that
\[
A_x = \sum_{i=1}^l \sum_{j=1}^l w_i w_j s_{ijx}.
\]
By the $\varepsilon$-balance property, $A_x \le \frac{1 + \varepsilon}{k}$ for all $x\in[k]$. Thus, for every $x\in[k]$,
\[
A_x = 1 -\sum_{y \neq x} A_y \ge \frac{1-\varepsilon (k-1)}{k}.
\]
On the other hand, since $S$ is intersecting,
\[
A_x 
\le \sum_{i=1}^l \sum_{j=1}^l w_i w_j \I[x \in F_i \cap F_j]
= \left( \sum_{i=1}^l w_i \I[x \in F_i] \right)^2
.\]
We also have
\[
\sum_{x \in [k]} \sum_{i=1}^l w_i \I[x \in F_i] 
= \sum_{i=1}^l w_i \sum_{x \in [k]} \I[x \in F_i]
= \sum_{i=1}^l w_i |F_i| 
= \ar(S)
.\]
Therefore, there exists $p'\in[k]$ such that
\[
\sum_{i=1}^l w_i \I[p' \in F_i] \le \frac{\ar(S)}{k}.
\]
For this $p'$, we get
\[
\frac{1-\varepsilon (k-1)}{k} \le A_{p'} \le \left( \frac{\ar(S)}{k} \right)^2.
\]
This implies that
\[
\ar(S)^2 \ge k (1-\varepsilon (k-1)).
\]

This concludes the proof of Theorem~\ref{t-bis-rank-lower}.

\subsection{Upper bound}

We now prove Theorem~\ref{t-bis-rank-upper}, following the plan sketched in Subsection~\ref{ss-res-bis}:
we build a good BIS for special values of $k$, and then extend it to arbitrary $k$
via the transition lemma, with only a small increase in rank.

\textbf{Explicit construction for special values of $k$.} For any prime power $q$, we can construct the projective plane over the finite field $\mathbb{F}_q$~\cite{Kalai2020Symmetric}. It is a symmetric intersecting family, and therefore it can be transformed into a BIS. See the proof of Theorem~\ref{t-formations-relation} for more details. This projective plane has $k = q^2 + q + 1$ elements and $k$ sets, and all sets have size $q+1$. Thus, we get the inequality below:
\[
    \mr(q^2 + q + 1) \le q+1.
\]

\textbf{Transition for arbitrary $k$.} In the following lemma, we effectively prove that
\[
    \mr(k+d) \le \max(\mr(k)+1,d).
\]
We also guarantee some additional properties of the constructed BIS.

\begin{lemma}
\label{l-system-solution-step}
Let $S = (\cF, w, s)$ be a balanced intersecting system on $k$ elements, let $l=|\cF|$, and let $r = \mr(S)$.

For any $d \in \N$, there exists a balanced intersecting system $S'$ on $k'=k+d$ elements with no more than $l + d$ sets such that
\[
    \mr(S') \le \max(r+1,d).
\]
Moreover, $S'$ can be computed from $S$ in time $O((l + d)^2 (k + d))$.
\end{lemma}

\begin{proof}

Denote $X = [k]$, $Y = \{k+1, \dots, k+d \}$, and let the new system be $S' = (\cF', w', s')$. We split the proof into two steps.

\textbf{Step 1.} First, we build an auxiliary family $\cF'' = (F_1'',\dots,F_{l''}'')$ and a weight vector $w'' \in [0, 1]^{l''}$ with the following properties:
\begin{enumerate}
    \item For every $i \in [l'']$, there exists $\phi(i) \in [l]$ and $y \in Y$ such that
    \[
        F_i'' = F_{\phi(i)} \cup \{ y \}.
    \]
    \item 
    For any $y \in Y$, the total weight of the sets $F'' \in \cF''$ that contain $y$ is equal to $1 / d$, i.e.\
    \[
        \forall y \in Y: \sum_{j} w''_j \I [y \in F_j''] = \frac{1}{d}.
    \]
    \item 
    Similarly, for sets $F \in \cF$:
    \[
        \forall i \in [l]: \sum_{j} w''_j \I [F_j'' \cap X = F_i] = w_i.
    \]
    \item $l'' \le l + d - 1$.
\end{enumerate}

Such a family $\cF''$ can be constructed by the two-pointers method as follows. Divide the segment $[0, 1]$ into consecutive subsegments of lengths $w_i$, and also divide the same segment $[0, 1]$ into consecutive subsegments of lengths $\frac{1}{d}$. Now consider the common refinement of these two partitions. For each subsegment of the refinement, say of length $t_i$, there is exactly one subsegment of the first partition and exactly one subsegment of the second partition that contain it. Suppose their indices are $i_x$ and $i_y$, respectively. Then we set
$F''_i = F_{i_x} \cup \{k + i_y\}$ and $w''_i = t_i$.
The number of subsegments in the common refinement is at most $l+d-1$, and the construction satisfies the four required properties by definition.

\textbf{Step 2.} Now, we define the resulting system $S' = (\cF', w', s')$ as follows:
\begin{itemize}
    \item $l' = l'' + 1$ and $\cF' = (F_1'', \dots, F_{l''}'' , Y)$.
    \item For any $i \neq l'$, define
    \[
        w'_i = w_i'' \cdot \sqrt{\frac{k}{k + d}},
    \]
    and set
    \[
        w'_{l'} = 1 - \sqrt{\frac{k}{k + d}}.
    \]
    \item \[
    s'_{ijx} =
    \begin{cases}
      s_{\phi(i) \phi(j) x} & \text{if } i \neq l', j  \neq l', x \in X, \\
      1 & \text{if } i \neq l', j = l', x \in F'_i \cap Y, \\
      1 & \text{if } i = l', j \neq l', x \in F'_j \cap Y,  \\
      \frac{1}{d} & \text{if } i = l', j = l', x \in Y, \\
      0 & \text{otherwise}. \\
    \end{cases}
    \]
\end{itemize}

It is straightforward to check that $S'$ is intersecting. Indeed, if $i,j\neq l'$, then positive values of $s'_{ijx}$ come from the original intersecting system $S$. If one of $i,j$ is equal to $l'$, then the only positive value is assigned to the unique point in the intersection with $Y$. Finally, if $i=j=l'$, then all positive values lie in $Y$.

It remains to check that $S'$ is balanced.

\begin{enumerate}
    \item Let $x \in X$.
    Using property 3 of $(\cF'', w'')$, we get:
    \[
        A_x(S') 
        = \sum_{i = 1}^{l'} \sum_{j = 1}^{l'}w'_i \cdot w'_j \cdot s'_{ijx} 
        = \frac{k}{k+d} \sum_{i = 1}^{l'-1} \sum_{j = 1}^{l'-1} w''_i \cdot w''_j \cdot s_{\phi(i) \phi(j) x}
        =
    \] \[
        = \frac{k}{k+d} \sum_{i = 1}^{l} \sum_{j = 1}^{l} w_i \cdot w_j \cdot s_{i j x}
        = \frac{k}{k+d} \cdot \frac{1}{k} 
        = \frac{1}{k + d}
        ;
    \]

    \item Let $x \in Y$.
    Using property 2 of $(\cF'', w'')$, we get:
    \[
        \sum_{i = 1}^{l'-1} w'_i  \cdot s'_{l'ix} 
        = \sqrt{\frac{k}{k + d}} \cdot\sum_{i = 1}^{l'-1} w''_i  \cdot \I(x \in F'_i)
        = \frac{1}{d} \sqrt{\frac{k}{k + d}}
        ;
    \]
    and therefore
    \[
        A_x(S')
        = \sum_{i = 1}^{l'} \sum_{j = 1}^{l'}w'_i \cdot w'_j \cdot s'_{ijx} 
        = 2 \cdot w'_{l'} \cdot \sum_{i = 1}^{l'-1} w'_i  \cdot s'_{l'ix} + (w'_{l'})^2 \cdot \frac{1}{d}
    \] \[
        = 2 \cdot \left( 1 - \sqrt{\frac{k}{k + d}} \right) \cdot \frac{1}{d} \cdot \sqrt{\frac{k}{k + d}} + \left( 1 - \sqrt{\frac{k}{k + d}} \right) ^ 2 \cdot \frac{1}{d} 
        = \frac{1}{k + d}.
    \]
\end{enumerate}
Thus $S'$ is balanced.

Finally, the rank of every set $F_i''$ is at most $r+1$, while the special set $Y$ has size $d$. Therefore,
\[
    \mr(S') \le \max(r+1,d).
\]
The number of sets is $l'=l''+1\le l+d$. The claimed running time follows directly from the construction of $\cF'$ and the array $s'$.
\end{proof}

\textbf{The final construction.} The resulting BIS is constructed as follows. Given $k$, compute the maximum prime power $q$ such that $q^2+q+1\le k$. Starting from the projective plane on $q^2+q+1$ elements, we repeatedly apply the construction from Lemma~\ref{l-system-solution-step} until we obtain a BIS on $k$ elements. At each application of the lemma, the rank increases by at most $1$, provided the number of new elements added at that step is at most the current rank plus $1$. The rest of this section analyzes the resulting procedure.

Define
\[
    w(r) := r^2 + r + 1
\]
and
\[
    Q(r) = \max \{q \; | \; q \le r , q \text{ is a prime power}\}.
\]
Recall that $D(r) = r - Q(r)$.

It is easy to see that the largest $r$ such that $w(r) \le k$ is
\[
    r'(k) = \floor{\sqrt{k-3/4} - 1/2}.
\]

Since $w(r + 1) - w(r) = 2r + 2$, we can get
\[
    \mr(w(r + 1)) \le \mr(w(r)) + 2
\]
by applying Lemma~\ref{l-system-solution-step} twice. Iterating this inequality gives
\[
\mr(w(r')) 
\le \mr(w(Q(r'))) + 2(r' - Q(r'))
\le Q(r') + 1 + 2 D(r')
=
r' + D(r') + 1.
\]

By maximality of $r'$, we have $w(r') \le k < w(r' + 1)$. Hence, using Lemma~\ref{l-system-solution-step} at most twice more, we obtain
\[
    \mr(k) \le \mr(w(r')) + 2 \le r' + D(r') + 3.
\]

Thus, we have constructed a BIS $S$ on $k$ elements with $\mr(S) \le r' + D(r') + 3$. After each application of the lemma, the number of sets in the BIS is at most the number of elements, so $S$ contains no more than $k$ sets. Each step can be computed in $O(k^3)$ time, and there are at most $O(k)$ steps. Therefore, the BIS $S$ can be computed in time $O(k^4)$.

By the prime gap theorem~\cite{Baker2001PrimeGap}, we know that $D(r) \le r^{0.525}$ for all sufficiently large $r$, and hence $D(r)=o_r(r)$. Therefore,
\[
    \mr(S) \le r'(k)+D(r'(k))+3=\sqrt k(1+o_k(1)).
\]

This concludes the proof of Theorem~\ref{t-bis-rank-upper}.

\section{Proofs of main theorems}
\label{s-main-proofs}

Here we prove Theorems~\ref{t-main-fixedn} and~\ref{t-main-bign}. Most of the statements follow directly from the results proved above.

\begin{itemize}
    \item \textbf{Theorem~\ref{t-main-fixedn}, upper bound.} We use the DBISP algorithm from Theorem~\ref{t-algo-deg} with a balanced intersecting system $S$ such that $\ar(S) = \ar(k)$, and any proper threshold $d_H$.

    Let us give a more precise analysis of the probabilistic guarantees. Denote by $f_G$ the partition produced by our algorithm on a graph $G$.
    Define $\alpha(n, m)$ as the smallest nonnegative number\footnote{The minimum exists since the number of graphs on $n$ vertices is finite, and the random variable $f_G$ is discrete.} such that, for every graph $G$ with $n$ vertices and $m$ edges,
    \[
        \bP[\arf(f_G) \le \ar(k) + \alpha(n, m)] \ge 2/3.
    \]
    By Theorem~\ref{t-algo-deg}, for every constant $\varepsilon > 0$ and all sufficiently large $n, m$, we have
    \[
        \bP[\arf(f_G) \le \ar(k) + \varepsilon] \ge 2/3
    \]
    for every graph $G$ with $n$ vertices and $m$ edges. Hence $\alpha(n, m) \le \varepsilon$ for all sufficiently large $n$ and $m$, and therefore 
    $\alpha(n, m)=o(1)$.

    Similarly, define $\beta(n, m)$ so that, for every graph $G$ with $n$ vertices and $m$ edges,
    \[
        \bP[\ib(f_G) \le 1 + \beta(n, m)] \ge 2/3.
    \]
    Again, Theorem~\ref{t-algo-deg} implies that $\beta(n, m)=o(1)$. Finally, by the union bound,
    \[
        \bP[\arf(f_G) \le \ar(k) + \alpha(n, m) \text{ and } \ib(f_G) \le 1 + \beta(n, m)] \ge 1/3.
    \]
    Thus, for every graph $G$, there exists a partition satisfying the desired bounds.

    \item \textbf{Theorem~\ref{t-main-fixedn}, lower bound.} This follows directly from Theorem~\ref{t-lower-mixing-fixedn}. For completeness, we also give a proof using the simpler lower bound from Theorem~\ref{t-lower-complete}.

    We argue by contradiction. Suppose that, for every $n$, there is a partition of $K_n$ such that
    \[
        \arf(f_n) \le \ar(k)-\varepsilon+o_n(1),
        \qquad
        \ib(f_n) \le 1+o_n(1).
    \]
    By Theorem~\ref{t-lower-complete}, there exists a sequence of intersecting systems
    \[
        \{S^{(n)} \}_{n=1}^\infty
        =
        \{(\cF^{(n)}, w^{(n)}, s^{(n)})\}_{n=1}^\infty
    \]
    such that
    \[
        \ar(S^{(n)}) \le \ar(k) - \varepsilon + o_n(1),
        \qquad
        \ib(S^{(n)}) \le 1 + o_n(1).
    \]
    By Lemma~\ref{l-system-limits}, this gives a balanced intersecting system $S^*$ with $\ar(S^*) \le \ar(k) - \varepsilon$, a contradiction.

    \item \textbf{Theorem~\ref{t-main-rank}.} This follows directly from Corollary~\ref{c-bis-rank}.

    \item \textbf{Theorem~\ref{t-main-bign}, upper bound.} We split the proof into two cases.
    
    The first (main) case: suppose that $m \ge n$.
    Consider a BIS $S$ such that $\ar(S) = \ar(k)$. Then we run the DBISP algorithm with this system $S$ and an integer $d_H$ satisfying
    \[
        d_H=\omega\left(\frac{m k}{n}\right)
        \qquad\text{and}\qquad
        d_H=o\left(\frac{m}{k^3\log k}\right).
    \]
    Such a choice is possible because 
    \[
        \frac{m k}{n} = o \left( \frac{m}{k^3\log k} \right)
        \qquad\text{and}\qquad
        \frac{m}{k^3\log k} = \omega(1).
    \]
    By Theorem~\ref{t-algo-deg}, we get a partition $f$ such that
    \[
        \arf(f)
        \le \ar(k)+o_P(1),
    \]
    and
    \[
        \ib(f) \le 1+o_P(1/ k).
    \]
    We pass from $o_P$ to deterministic $o$ bounds using the same argument as in the proof of the first part of Theorem~\ref{t-main-fixedn}. Finally, we use Theorem~\ref{t-bis-rank-upper} to get the bound $\ar(k) + o(1) \le \sqrt{k} (1 + o(1))$ and finish the proof.

    The second case: suppose that $m < n$. Since the replication factor of any vertex is at most its degree, for any edge partition $f$ we have
    \[
        \arf(f) \le \overline{d} = O(1) = o(\sqrt{k})=o(\ar(k)).
    \]
    We take any distribution of edges in which the sizes of any two parts differ by at most one. Then $\ib(f) \le \ceil{m/k}/(m/k) \le 1 + k/m = 1 + o(1/k)$, since $k^2 = o(m)$.

    \item \textbf{Theorem~\ref{t-main-bign}, lower bound.} This follows directly from Theorem~\ref{t-lower-mixing-bign}. Alternatively, we can prove it using the lower bound for complete graphs as follows.
    
    Suppose the contrary. Consider a partition of the complete graph $K_n$ with
    \[
        \arf(f) \le \sqrt{k}(1-\varepsilon+o(1/ k)),
        \qquad
        \ib(f) \le 1+o(1/ k).
    \]
    By Theorem~\ref{t-lower-complete}, there exists an intersecting system $S$ such that
    \[
        \ar(S) \le \sqrt{k}(1-\varepsilon+o(1/ k)),
        \qquad
        \ib(S) \le 1+o(1/k).
    \]
    But by Theorem~\ref{t-bis-rank-lower}, such a system satisfies
    \[
        \ar(S) \ge \sqrt{k(1-o(1))}.
    \]
    For sufficiently large $k$, this is a contradiction.

\end{itemize}  

\section{Relation between BIS, RIF and SIF}
\label{s-relation}
In this section, we prove Theorem~\ref{t-formations-relation}.

\begin{enumerate}
\item \textbf{Any symmetric intersecting family is balanceable.} \\
Let $\cF \subset 2^X$ be a symmetric intersecting family, and let $k = |X|$, $l=|\cF|$. Define
\[
    w_i = \frac{1}{l};
\]
\[
    s_{ijp} = \frac{\I [p \in F_i \cap F_j]}{|F_i \cap F_j|}.
\]
Since $\cF$ is intersecting, $s$ is well defined. Clearly, $(\cF, w, s)$ is an intersecting system. Since $\cF$ is symmetric, we have $A_x = A_y$ for all $x,y \in X$. Also, $\sum_{x\in X} A_x = 1$, and therefore $A_x = \frac{1}{k}$ for every $x \in X$. Thus, $(\cF,w,s)$ is balanced.

\item \textbf{There exists a uniform and balanceable family that is not regular.} \\
The proof idea is as follows: take a symmetric intersecting, and hence balanceable, family $\cF'$. If we enlarge every set in this family, then the resulting family remains balanceable. We can enlarge the sets unevenly so that the resulting family is no longer regular.

Let us give a formal proof. Take $X = \Z/(10 \Z)$. The family
\[
    \cF' = \{ \{x, x+1, x+2, x+5\} \; | \; x \in X \}
\]
is balanceable because it is symmetric and intersecting. Now define
\[
\cF = \{ \{x, x+1, x+2, x+5, x+6\} \; | \; x \in X, x \text{ is even}\} \cup
\] \[
\cup \{ \{x, x+1, x+2, x+5, x+7\} \; | \; x \in X, x \text{ is odd}\}.
\]
This family is obtained from $\cF'$ by enlarging each set. Therefore, the same weights and the same choice probabilities as for $\cF'$ still define a balanced intersecting system: all intersections only become larger, while the probabilities may still be supported on the old intersections. Hence $\cF$ is balanceable. However, it is not regular, because even elements have higher degree than odd elements.

\item \textbf{There exists a uniform, regular, and intersecting family that is not balanceable.} \\
The idea of an example is as follows.
Let $X = X_1 \sqcup X_2$.
Let $\cF_1$ be a uniform regular, but not intersecting, family on $X_1$.
Let $\cF_2$ be a uniform regular intersecting family on $X_2$.
Define the family $\cF = \cF_1 + \cF_2 = \{F_1 \cup F_2 \; | \; F_1 \in \cF_1, F_2 \in \cF_2\}$.
The family can be made regular, and any two of its sets intersect in some element of $X_2$.
However, we can make intersections between the sets in $\cF_1$ so rare that the family will not be balanceable.

Let us give a formal proof.

Let $X_1 = [2 k_1]$, where $k_1=10$. Let
\[
    \cF_1 = \{\{1,  \dots, k_1\},  \{k_1 +1, \dots, 2k_1\}\},
\]
i.e.\ $\cF_1$ consists of two disjoint sets of size $k_1$. Clearly, $\cF_1$ is regular and uniform.

Let $X_2 = \Z / (k_2 \Z)$, where $k_2 = 6$. We denote elements of $\Z / (k_2)$ as $\overline{x}$, $x \in \Z$, to avoid confusion with elements of $\Z$. Define
\[
    \cF_2 = \{\{\overline{1}+x, \overline{2}+x, \overline{4}+x\} \; | \; x \in \Z / (k_2) \}.
\]
The family $\cF_2$ is uniform, regular, and intersecting\footnote{The intersecting property can be checked by brute force. Alternatively, it follows from the fact that $A=\{\overline{1}, \overline{2}, \overline{4}\}$ is a difference cover in $\Z / (6)$, i.e.\ $A-A = \Z / (6)$. See~\cite{Kalai2020Symmetric} for more details.}.

Consider the family $\cF = \cF_1 + \cF_2$ on $X = X_1 \sqcup X_2$. It is intersecting because $\cF_2$ is intersecting. It is regular because every element lies in one half of all sets. It is uniform because both $\cF_1$ and $\cF_2$ are uniform.

We need to prove that $\cF$ is not balanceable. Suppose that $(\cF, w, s)$ is a balanced intersecting system. Let $w'_1$ be the total weight, with respect to $w$, of the sets containing $1$, and let $w'_2$ be the total weight of the sets containing $k_1 + 1$. We have $w'_1 + w'_2 = 1$, so without loss of generality $w'_1 \le \frac{1}{2}$. Then
\[
\sum_{x=1}^{k_1} A_x 
= \sum_{i=1}^{|\cF|} \sum_{j=1}^{|\cF|} w_i w_j \sum_{x=1}^{k_1} s_{ijx} 
\le \sum_{i=1}^{|\cF|} \sum_{j=1}^{|\cF|} w_i w_j \I [ \{1, \dots, k_1\} \cap F_i \cap F_j \neq \varnothing] \le 
\] \[
\le \left( \sum_{i=1}^{|\cF|} w_i \I[1 \in F_i] \right)^2 = (w'_1)^2 \le \frac{1}{4}.
\]
On the other hand, since the system is balanced,
\[
    \sum_{x=1}^{k_1}  A_x  = \frac{k_1}{2 k_1 + k_2} = \frac{10}{26} > \frac{1}{4}.
\]
This contradiction shows that $\cF$ is not balanceable.

\item \textbf{There exists a uniform, regular, balanceable and intersecting family that is not symmetric.} \\
Consider a non-symmetric projective plane $\cF$; for instance, one can take a Hughes plane~\cite{Hughes1957} of order $9$. By definition, $\cF$ is uniform, regular, and intersecting. Let us show that it is balanceable.

Let $q$ be the order of the projective plane. Then the number of points and lines is
\[
    k = q^2+q+1,
\]
each line has size $q+1$, and each point lies on $q+1$ lines. Define $w_i=1/k$ for every line $F_i$, and define
\[
    s_{ijx} = \frac{\I[x\in F_i\cap F_j]}{|F_i\cap F_j|}.
\]
This gives an intersecting system. For any point $x$, consider the sum
\[
    \sum_{i=1}^l\sum_{j=1}^l s_{ijx}.
\]
The diagonal terms $s_{iix}$ contribute
\[
    (q+1)\cdot \frac{1}{k^2(q+1)}=\frac{1}{k^2}.
\]
The off-diagonal terms come from ordered pairs of distinct lines through $x$. There are $q(q+1)$ such pairs, and each contributes $1/k^2$. Therefore,
\[
    A_x
    =
    \frac{1}{k^2}+\frac{q(q+1)}{k^2}
    =
    \frac{q^2+q+1}{k^2}
    =
    \frac{1}{k}.
\]
Thus the system is balanced, and so $\cF$ is balanceable.

\end{enumerate}

This concludes the proof of Theorem~\ref{t-formations-relation}.

\begin{remark}
    \label{r-relations-opt-unknown}
    Theorem~\ref{t-formations-relation} gives a complete understanding of how the different conditions on uniform intersecting families are related to each other. However, what can be said about the corresponding optimal ranks, that is, about the values $\ar(k)$, $\mr(k)$, $\rho_r(k)$, and $\rho_s(k)$?

    For $k = 2$, the family $\{\{1\}, \{1, 2\}\}$ with suitable weights gives $\ar(S) = 1 + 1/\sqrt{2}$ (see Figure \ref{f-bis-n2}). At the same time, the only symmetric or regular intersecting family is $\{\{1, 2\}\}$, which has average set size $2$, even if we do not require uniformity. Thus, the optimal \textit{average} rank for balanced systems is strictly smaller than for symmetric or regular families.
    
    For the \textit{maximum} set size, however, we do not know any value of $k$ for which such a difference is known.
\end{remark}

\section{Conclusion}
\label{s-final}

We split our conclusion into two parts: practical and theoretical.

\subsection{Practical insights}

We provide a theoretical study of a class of edge-partitioning algorithms used in distributed graph-processing frameworks. We give algorithms with the best known guarantees for an arbitrary number of parts $k$. More generally, we prove that the BISP algorithm with an optimal system provides optimal guarantees for fixed $k$. Thus, our approach is optimal within this class of algorithms.

BIS-based partitions are near-optimal on pseudo-random graphs, but real-world graphs are usually very different. We suggest the following interpretation of our lower bounds: to improve over the BIS approach, an algorithm must exploit specific properties of the input graph, such as a power-law degree distribution or the presence of dense communities. Another possible direction for improvement is to optimize metrics other than imbalance and replication factor, depending on the application, cluster configuration, and other practical constraints.

\subsection{Theoretical insights}

We study the edge-partitioning problem in regimes where the number of parts $k$ is small, either constant or slowly growing, and where almost perfect balance is required. We prove asymptotically tight bounds, up to a $1+o(1)$ factor, in these regimes and provide algorithms for finding such partitions. We see several possible ways to generalize, extend, and improve these results.
\begin{itemize}
    \item It may be possible to improve the known results about $\ar(k)$ and $\mr(k)$. We expect that finding an exact formula for every $k$ is a very hard problem, similarly to finding the optimal rank of SIFs and RIFs.
    Some easier but still interesting questions are the following: can one prove a bound $\ar(k)^2 - \ar(k) +1 \ge k$, analogous to the bound in~\cite{Furedi1981MaxDegree}? Is it true that $\mr(k) \le \sqrt{k} + O(1)$?
    The relation between these quantities and the optimal ranks of symmetric and regular intersecting families is also interesting; see Remark~\ref{r-relations-opt-unknown}.

    \item It would be interesting to understand the possible trade-offs between imbalance and replication factor. In particular, when $k$ grows, we found the optimal partition when we require $\ib(f) = 1 + o(1/k)$. What happens if this condition is relaxed, for example to 
    $\ib(f) = 1 + o(1)$?
    
    \item What happens in the regime where the number of parts $k$ is larger, for example when $k = \Theta(n^\alpha)$ for different values of $\alpha$?

    \item In this work, we focus on the best possible worst-case guarantees, that is, on the smallest $\alpha$ such that every graph has a balanced partition with $\arf(f) \le \alpha$. It would also be interesting to study the problem from the point of view of approximation algorithms: what is the best $\alpha$ for which one can efficiently build a partition with $\arf(f) \le \alpha \cdot OPT$, where $OPT$ is the replication factor of an optimal partition?
\end{itemize}

\bibliography{bib}

@article{Wegman1981KIndHash,
  title = {New hash functions and their use in authentication and set equality},
  volume = {22},
  ISSN = {0022-0000},
  DOI = {10.1016/0022-0000(81)90033-7},
  number = {3},
  journal = {Journal of Computer and System Sciences},
  publisher = {Elsevier BV},
  author = {Wegman,  Mark N. and Carter, J. Lawrence},
  year = {1981},
  month = jun,
  pages = {265–279}
}

@article{Janson2004,
  title = {Large deviations for sums of partly dependent random variables},
  volume = {24},
  ISSN = {1098-2418},
  DOI = {10.1002/rsa.20008},
  number = {3},
  journal = {Random Structures \& Algorithms},
  publisher = {Wiley},
  author = {Janson,  Svante},
  year = {2004},
  month = mar,
  pages = {234–248}
}

@article{Baker2001PrimeGap,
  title = {The Difference Between Consecutive Primes,  {II}},
  volume = {83},
  ISSN = {0024-6115},
  DOI = {10.1112/plms/83.3.532},
  number = {3},
  journal = {Proceedings of the London Mathematical Society},
  publisher = {Wiley},
  author = {Baker,  R. C. and Harman,  G. and Pintz,  J.},
  year = {2001},
  month = nov,
  pages = {532–562}
}

@incollection{thomason1987PseudoRandom,
  title={Pseudo-random graphs},
  author={Thomason, Andrew},
  booktitle={North-Holland Mathematics Studies},
  volume={144},
  pages={307--331},
  year={1987},
  publisher={Elsevier}
}

@incollection{krivelevich2006PseudoRandom,
  title={Pseudo-random graphs},
  author={Krivelevich, Michael and Sudakov, Benny},
  booktitle={More sets, graphs and numbers: A Salute to Vera Sos and Andr{\'a}s Hajnal},
  pages={199--262},
  year={2006},
  publisher={Springer}
}

@article{Furedi1981MaxDegree,
  title = {Maximum degree and fractional matchings in uniform hypergraphs},
  volume = {1},
  ISSN = {1439-6912},
  DOI = {10.1007/bf02579271},
  number = {2},
  journal = {Combinatorica},
  publisher = {Springer Science and Business Media LLC},
  author = {F\"{u}redi,  Zoltán},
  year = {1981},
  month = jun,
  pages = {155–162}
}

@article{Furedi1986twise,
  title = {Finite projective spaces and intersecting hypergraphs},
  volume = {6},
  ISSN = {1439-6912},
  DOI = {10.1007/bf02579260},
  number = {4},
  journal = {Combinatorica},
  publisher = {Springer Science and Business Media LLC},
  author = {Frankl,  P. and F\"{u}redi,  Z.},
  year = {1986},
  month = dec,
  pages = {335–354}
}

@article{Cameron1989Cardinality,
    author = {Cameron, P.J. and Frankl, P. and Kantor, W.M.},
    title = {Intersecting Families of Finite Sets and Fixed-point-Free 2-Elements},
    journal = {European Journal of Combinatorics},
    volume={10}, number={2},
    year = {1989},
    pages = {149-160}
}

@article{Furedi1990CoveringPairs,
  title = {Covering pairs by $q^2+q+1$ sets},
  volume = {54},
  ISSN = {0097-3165},
  DOI = {10.1016/0097-3165(90)90034-t},
  number = {2},
  journal = {Journal of Combinatorial Theory,  Series A},
  publisher = {Elsevier BV},
  author = {F\"{u}redi,  Z},
  year = {1990},
  month = jul,
  pages = {248–271}
}

@article{Kupavskii2019Regular,
  title = {Regular intersecting families},
  volume = {270},
  ISSN = {0166-218X},
  DOI = {10.1016/j.dam.2019.07.009},
  journal = {Discrete Applied Mathematics},
  publisher = {Elsevier BV},
  author = {Ihringer,  Ferdinand and Kupavskii,  Andrey},
  year = {2019},
  month = nov,
  pages = {142–152}
}

@article{Kalai2020Symmetric,
  title = {On symmetric intersecting families},
  volume = {86},
  ISSN = {0195-6698},
  DOI = {10.1016/j.ejc.2020.103094},
  journal = {European Journal of Combinatorics},
  publisher = {Elsevier BV},
  author = {Ellis,  David and Kalai,  Gil and Narayanan,  Bhargav},
  year = {2020},
  month = may,
  pages = {103094}
}

@article{Singers38OnFPP,
    author = {Singer, J.},
    title = {A theorem in finite projective geometry and some applications to number theory},
    journal = {Trans. Amer. Math. Soc},
    year = {1938},
    pages = {377-385}
}

@article{Bruck1949FPPNonExistence,
  title = {The Nonexistence of Certain Finite Projective Planes},
  volume = {1},
  ISSN = {1496-4279},
  DOI = {10.4153/cjm-1949-009-2},
  number = {1},
  journal = {Canadian Journal of Mathematics},
  publisher = {Canadian Mathematical Society},
  author = {Bruck,  R. H. and Ryser,  H. J.},
  year = {1949},
  month = feb,
  pages = {88–93}
}

@article{Hughes1957,
  title = {A Class of Non-Desarguesian Projective Planes},
  volume = {9},
  ISSN = {1496-4279},
  DOI = {10.4153/cjm-1957-045-0},
  journal = {Canadian Journal of Mathematics},
  publisher = {Canadian Mathematical Society},
  author = {Hughes,  D. R.},
  year = {1957},
  pages = {378–388}
}

@article{Lam1989NoFPP10,
  title = {The Non-Existence of Finite Projective Planes of Order 10},
  volume = {41},
  ISSN = {1496-4279},
  DOI = {10.4153/cjm-1989-049-4},
  number = {6},
  journal = {Canadian Journal of Mathematics},
  publisher = {Canadian Mathematical Society},
  author = {Lam,  C. W. H. and Thiel,  L. and Swiercz,  S.},
  year = {1989},
  month = Dec,
  pages = {1117–1123}
}

@article{Lam1991NoFPP10Exp,
  title = {The Search for a Finite Projective Plane of Order 10},
  volume = {98},
  ISSN = {1930-0972},
  DOI = {10.1080/00029890.1991.12000759},
  number = {4},
  journal = {The American Mathematical Monthly},
  publisher = {Informa UK Limited},
  author = {Lam,  C. W. H.},
  year = {1991},
  month = apr,
  pages = {305–318}
}

@article{Akiyama2019TheNoFPP12,
  title={The nonexistence of projective planes of order 12 with a collineation group of order 9},
  author={Kenzi Akiyama and Chihiro Suetake and Masaki Tanaka},
  journal={Australas. J Comb.},
  year={2019},
  volume={74},
  pages={112-160},
}

@article{RenyiBasicEstimation1949,
    author = {R{\'e}dei, L. and R{\'e}nyi, A.},
    volume = {66},
    title = {On the representation of the numbers 1, 2, ..., $N$ by means of differences},
    journal = {Mat. Sb.},
    year = {1949},
    pages = {385-389}
}

@inproceedings{malewicz2010pregel,
  title={Pregel: a system for large-scale graph processing},
  author={Malewicz, Grzegorz and Austern, Matthew H and Bik, Aart JC and Dehnert, James C and Horn, Ilan and Leiser, Naty and Czajkowski, Grzegorz},
  booktitle={Proceedings of the 2010 ACM SIGMOD International Conference on Management of data},
  pages={135--146},
  year={2010}
}

@inproceedings{low2010graphlab,
    author = {Low, Yucheng and Gonzalez, Joseph and Kyrola, Aapo and Bickson, Danny and Guestrin, Carlos and Hellerstein, Joseph},
    title = {GraphLab: a new framework for parallel machine learning},
    year = {2010},
    isbn = {9780974903965},
    publisher = {AUAI Press},
    address = {Arlington, Virginia, USA},
    
    booktitle = {Proceedings of the Twenty-Sixth Conference on Uncertainty in Artificial Intelligence},
    pages = {340–349},
    numpages = {10},
    location = {Catalina Island, CA},
    series = {UAI'10}
}

@inproceedings{gonzalez2012powergraph,
  title={{PowerGraph}: Distributed Graph-Parallel Computation on Natural Graphs},
  author={Gonzalez, Joseph E and Low, Yucheng and Gu, Haijie and Bickson, Danny and Guestrin, Carlos},
  booktitle={10th USENIX symposium on operating systems design and implementation (OSDI 12)},
  pages={17--30},
  year={2012}
}

@inproceedings{gonzalez2014graphx,
  title={{GraphX}: Graph Processing in a Distributed Dataflow Framework},
  author={Gonzalez, Joseph E and Xin, Reynold S and Dave, Ankur and Crankshaw, Daniel and Franklin, Michael J and Stoica, Ion},
  booktitle={11th USENIX symposium on operating systems design and implementation (OSDI 14)},
  pages={599--613},
  year={2014}
}

@inbook{Kumar2017CostModel,
  title = {Cost Model for Pregel on GraphX},
  ISBN = {9783319669175},
  ISSN = {1611-3349},
  DOI = {10.1007/978-3-319-66917-5_11},
  booktitle = {Advances in Databases and Information Systems},
  publisher = {Springer International Publishing},
  author = {Kumar,  Rohit and Abelló,  Alberto and Calders,  Toon},
  year = {2017},
  pages = {153–166}
}

@inproceedings{Zaharia10Spark,
    author = {Zaharia, Matei and Chowdhury, Mosharaf and Franklin, Michael J. and Shenker, Scott and Stoica, Ion},
    title = {Spark: cluster computing with working sets},
    year = {2010},
    publisher = {USENIX Association},
    address = {USA},
    booktitle = {Proceedings of the 2nd USENIX Conference on Hot Topics in Cloud Computing},
    pages = {10},
    numpages = {1},
    location = {Boston, MA},
    series = {HotCloud'10}
}

@inproceedings{Roy2013xstream,
  series = {SOSP ’13},
  title = {X-Stream: edge-centric graph processing using streaming partitions},
  DOI = {10.1145/2517349.2522740},
  booktitle = {Proceedings of the Twenty-Fourth ACM Symposium on Operating Systems Principles},
  publisher = {ACM},
  author = {Roy,  Amitabha and Mihailovic,  Ivo and Zwaenepoel,  Willy},
  year = {2013},
  month = nov,
  pages = {472–488},
  collection = {SOSP ’13}
}

@article{chen2019PowerLyra,
  title={Powerlyra: Differentiated graph computation and partitioning on skewed graphs},
  author={Chen, Rong and Shi, Jiaxin and Chen, Yanzhe and Zang, Binyu and Guan, Haibing and Chen, Haibo},
  journal={ACM Transactions on Parallel Computing (TOPC)},
  volume={5},
  number={3},
  pages={1--39},
  year={2019},
  publisher={ACM New York, NY, USA}
}

@inproceedings{Zhao2014LightGraph,
  title = {LightGraph: Lighten Communication in Distributed Graph-Parallel Processing},
  DOI = {10.1109/bigdata.congress.2014.106},
  booktitle = {2014 IEEE International Congress on Big Data},
  publisher = {IEEE},
  author = {Zhao,  Yue and Yoshigoe,  Kenji and Xie,  Mengjun and Zhou,  Suijian and Seker,  Remzi and Bian,  Jiang},
  year = {2014},
  month = jun,
  pages = {717–724}
}

@article{Li2019TopoX,
author = {Li, Dongsheng and Zhang, Yiming and Wang, Jinyan and Tan, Kian-Lee},
title = {TopoX: topology refactorization for efficient graph partitioning and processing},
year = {2019},
issue_date = {April 2019},
publisher = {VLDB Endowment},
volume = {12},
number = {8},
issn = {2150-8097},
doi = {10.14778/3324301.3324306},
journal = {Proc. VLDB Endow.},
month = apr,
pages = {891–905},
numpages = {15}
}

@article{Coimbra2021ReviewGraphProcessing,
  title = {An analysis of the graph processing landscape},
  volume = {8},
  ISSN = {2196-1115},
  DOI = {10.1186/s40537-021-00443-9},
  number = {1},
  journal = {Journal of Big Data},
  publisher = {Springer Science and Business Media LLC},
  author = {Coimbra,  Miguel E. and Francisco,  Alexandre P. and Veiga,  Luís},
  year = {2021},
  month = Apr 
}

@article{McCune2015ReviewTLAV,
  title = {Thinking Like a Vertex: A Survey of Vertex-Centric Frameworks for Large-Scale Distributed Graph Processing},
  volume = {48},
  ISSN = {1557-7341},
  DOI = {10.1145/2818185},
  number = {2},
  journal = {ACM Computing Surveys},
  publisher = {Association for Computing Machinery (ACM)},
  author = {McCune,  Robert Ryan and Weninger,  Tim and Madey,  Greg},
  year = {2015},
  month = Oct,
  pages = {1–39}
}

@article{atalyrek2010OriginalGrid,
  title = {On Two-Dimensional Sparse Matrix Partitioning: Models,  Methods,  and a Recipe},
  volume = {32},
  ISSN = {1095-7197},
  DOI = {10.1137/080737770},
  number = {2},
  journal = {SIAM Journal on Scientific Computing},
  publisher = {Society for Industrial & Applied Mathematics (SIAM)},
  author = {{\c{C}}ataly{\"u}rek, {\"U}mit V. and Aykanat, Cevdet and U{\c{c}}ar, Bora},
  year = {2010},
  month = Jan,
  pages = {656–683}
}

@article{xie2014DegreeBased,
  title={Distributed power-law graph computing: Theoretical and empirical analysis},
  author={Xie, Cong and Yan, Ling and Li, Wu-Jun and Zhang, Zhihua},
  journal={Advances in neural information processing systems},
  volume={27},
  year={2014}
}

@inproceedings{Qu2023LikeFPPHolistic,
  title = {Optimizing Graph Partition by Optimal Vertex-Cut: A Holistic Approach},
  DOI = {10.1109/icde55515.2023.00083},
  booktitle = {2023 IEEE 39th International Conference on Data Engineering (ICDE)},
  publisher = {IEEE},
  author = {Qu,  Wenwen and Zhang,  Weixi and Cheng,  Ji and Zhang,  Chaorui and Han,  Wei and Bai,  Bo and Zhang,  Chen Jason and He,  Liang and Wang,  Xiaoling},
  year = {2023},
  month = apr,
  pages = {1019–1031}
}

@inproceedings{Zhang2023LikeFPPCombinatorial,
  title = {A Mixed-State Streaming Edge Partitioning based on Combinatorial Design},
  DOI = {10.1109/icdm58522.2023.00096},
  booktitle = {2023 IEEE International Conference on Data Mining (ICDM)},
  publisher = {IEEE},
  author = {Zhang,  Zhenyu and Qu,  Wenwen and Zhang,  Weixi and Shang,  Junlin and Wang,  Xiaoling},
  year = {2023},
  month = dec,
  pages = {868–877}
}

@inproceedings{bourse2014BGEP,
  title={Balanced graph edge partition},
  author={Bourse, Florian and Lelarge, Marc and Vojnovic, Milan},
  booktitle={Proceedings of the 20th ACM SIGKDD international conference on Knowledge discovery and data mining},
  pages={1456--1465},
  year={2014}
}

@inproceedings{petroni2015hdrf,
  title={Hdrf: Stream-based partitioning for power-law graphs},
  author={Petroni, Fabio and Querzoni, Leonardo and Daudjee, Khuzaima and Kamali, Shahin and Iacoboni, Giorgio},
  booktitle={Proceedings of the 24th ACM international on conference on information and knowledge management},
  pages={243--252},
  year={2015}
}

@inproceedings{zhang2017AlgoNe,
  title={Graph edge partitioning via neighborhood heuristic},
  author={Zhang, Chenzi and Wei, Fan and Liu, Qin and Tang, Zhihao Gavin and Li, Zhenguo},
  booktitle={Proceedings of the 23rd ACM SIGKDD International Conference on Knowledge Discovery and Data Mining},
  pages={605--614},
  year={2017}
}

@inproceedings{jain2013GraphBuilder,
  series = {SIGMOD/PODS’13},
  title = {GraphBuilder: scalable graph ETL framework},
  DOI = {10.1145/2484425.2484429},
  booktitle = {First International Workshop on Graph Data Management Experiences and Systems},
  publisher = {ACM},
  author = {Jain,  Nilesh and Liao,  Guangdeng and Willke,  Theodore L.},
  year = {2013},
  month = jun,
  pages = {1–6},
  collection = {SIGMOD/PODS’13}
}

@inproceedings{Mayer2018Adwise,
  title = {ADWISE: Adaptive Window-Based Streaming Edge Partitioning for High-Speed Graph Processing},
  DOI = {10.1109/icdcs.2018.00072},
  booktitle = {2018 IEEE 38th International Conference on Distributed Computing Systems (ICDCS)},
  publisher = {IEEE},
  author = {Mayer,  Christian and Mayer,  Ruben and Tariq,  Muhammad Adnan and Geppert,  Heiko and Laich,  Larissa and Rieger,  Lukas and Rothermel,  Kurt},
  year = {2018},
  month = jul,
  pages = {685–695}
}

@article{Hanai2019,
  title = {Distributed edge partitioning for trillion-edge graphs},
  volume = {12},
  ISSN = {2150-8097},
  DOI = {10.14778/3358701.3358706},
  number = {13},
  journal = {Proceedings of the VLDB Endowment},
  publisher = {Association for Computing Machinery (ACM)},
  author = {Hanai,  Masatoshi and Suzumura,  Toyotaro and Tan,  Wen Jun and Liu,  Elvis and Theodoropoulos,  Georgios and Cai,  Wentong},
  year = {2019},
  month = sep,
  pages = {2379–2392}
}

@inbook{Rahimian2014,
  title = {Distributed Vertex-Cut Partitioning},
  ISBN = {9783662433522},
  ISSN = {1611-3349},
  DOI = {10.1007/978-3-662-43352-2_15},
  booktitle = {Distributed Applications and Interoperable Systems},
  publisher = {Springer Berlin Heidelberg},
  author = {Rahimian,  Fatemeh and Payberah,  Amir H. and Girdzijauskas,  Sarunas and Haridi,  Seif},
  year = {2014},
  pages = {186–200}
}

@inproceedings{Mayer2021,
  series = {SIGMOD/PODS ’21},
  title = {Hybrid Edge Partitioner: Partitioning Large Power-Law Graphs under Memory Constraints},
  DOI = {10.1145/3448016.3457300},
  booktitle = {Proceedings of the 2021 International Conference on Management of Data},
  publisher = {ACM},
  author = {Mayer,  Ruben and Jacobsen,  Hans-Arno},
  year = {2021},
  month = jun,
  pages = {1289–1302},
  collection = {SIGMOD/PODS ’21}
}

@article{Verma2017Compare,
  title = {An experimental comparison of partitioning strategies in distributed graph processing},
  volume = {10},
  ISSN = {2150-8097},
  DOI = {10.14778/3055540.3055543},
  number = {5},
  journal = {Proceedings of the VLDB Endowment},
  publisher = {Association for Computing Machinery (ACM)},
  author = {Verma,  Shiv and Leslie,  Luke M. and Shin,  Yosub and Gupta,  Indranil},
  year = {2017},
  month = jan,
  pages = {493–504}
}

@article{abbas2018streaming,
  title={Streaming graph partitioning: an experimental study},
  author={Abbas, Zainab and Kalavri, Vasiliki and Carbone, Paris and Vlassov, Vladimir},
  journal={Proceedings of the VLDB Endowment},
  volume={11},
  number={11},
  pages={1590--1603},
  year={2018}
}

@inproceedings{pacaci2019experimental,
  title={Experimental analysis of streaming algorithms for graph partitioning},
  author={Pacaci, Anil and {\"O}zsu, M. Tamer},
  booktitle={Proceedings of the 2019 International Conference on Management of Data (SIGMOD)},
  pages={1375--1392},
  year={2019}
}

\end{document}